\documentclass[runningheads]{llncs}
\usepackage{graphicx}
\usepackage[utf8]{inputenc}
\usepackage{johofooter}
\usepackage{amssymb}
\usepackage{mathrsfs}
\usepackage{amsmath}
\usepackage{bussproofs}
\usepackage{tikz}
\usetikzlibrary{positioning, calc, shapes, patterns}
\tikzstyle{input}=[draw, rounded corners, rectangle,fill=colIn,inner ysep=2pt]
\tikzstyle{instance}=[draw, rounded corners, rectangle,fill=colInst,inner ysep=2pt]
\tikzstyle{lemma}=[draw, rounded corners, rectangle,fill=colLemma,inner ysep=2pt]
\tikzstyle{resolvent}=[draw, rounded corners, rectangle,fill=white,inner ysep=2pt]
\tikzstyle{interpol}=[rectangle, xshift=-0.3cm]
\usepackage{todonotes}
\usepackage{xspace}
\usepackage{enumitem}
\usepackage[most]{tcolorbox}
\usepackage{hyperref}
\tcbset{on line, colframe=black, size=fbox, boxsep=2pt, boxrule=0.3pt, arc=0.75mm}

\newtcbox{\myboxIn}[1][colIn]{colback=#1}
\newtcbox{\myboxInst}[1][colInst]{colback=#1}
\newtcbox{\myboxLemma}[1][colLemma]{colback=#1}
\newtcbox{\myboxRes}[1][white]{colback=#1}

\newcommand\T{\ensuremath{\mathcal{T}}\xspace}
\newcommand\symbset{\ensuremath{\mathop{\mathit{symb}}}}

\newcommand\project[2]{{#1}\downharpoonright{#2}}
\newcommand\compl[1]{\ensuremath{#1^c}}

\newcommand{\minproject}[2]{\ensuremath{{#1}\downharpoonright^{-}{#2}}}
\newcommand\set[1]{\ensuremath{\{#1\}}}

\newcommand{\hd}{\ensuremath{\mathit{hd}}}
\newcommand{\FreeVars}{\ensuremath{\mathit{FreeVars}}}
\newcommand{\varV}{\ensuremath{\mathcal{V}}}

\newcommand{\tree}{\ensuremath{\mathit{st}}}

\newcommand{\itpcolour}{\ensuremath{\mathit{colour}}}
\newcommand{\partitions}{\ensuremath{\mathit{partitions}}}
\newcommand{\flatten}{\ensuremath{\mathit{flatten}}}
\newcommand{\FlatEQ}{\ensuremath{\mathit{FlatEQ}}}

\newcommand{\AuxEQ}{\ensuremath{\mathit{AuxEQ}}}
\newcommand{\Supported}{\ensuremath{\mathit{Supported}}}

\newcommand{\ie}{i.\,e.}
\newcommand{\eg}{e.\,g.}

\newcommand\TODO[1]{{\textcolor{red}{TODO: #1}}}

\newcommand{\ignore}[1]{}

\colorlet{colLemma}	{blue!20}
\colorlet{colInst}	{yellow!40}
\colorlet{colIn}	{green!30}

\hypersetup{colorlinks=true,linkcolor=blue}

\begin{document}
\title{Choose your Colour: Tree Interpolation for Quantified Formulas in SMT}

\titlerunning{Tree Interpolation for Quantified Formulas}
%
\author{Elisabeth Henkel\inst{1}\orcidID{0000-0003-3844-8292} \and
Jochen Hoenicke\inst{2}\orcidID{0000-0002-6314-1041} \and
Tanja Schindler\inst{3}\orcidID{0000-0002-7462-8445}}
\authorrunning{Henkel et al.}
%
\institute{University of Freiburg, 
\email{henkele@informatik.uni-freiburg.de} \and
Certora, \email{jochen@certora.com} \and
University of Liège,
\email{tanja.schindler@uliege.be}
}
\maketitle              
\johofooter{lncs}{\emph{CADE-29}}{}{}
\begin{abstract}
We present a generic tree-interpolation algorithm in the SMT context with quantifiers.
The algorithm takes a proof of unsatisfiability using resolution and quantifier instantiation and computes
interpolants (which may contain quantifiers). 
Arbitrary SMT theories are supported, as long as each theory itself supports tree interpolation for its lemmas.
In particular, we show this for the theory combination of equality with uninterpreted functions and linear arithmetic.
The interpolants can be tweaked by virtually assigning each literal in the proof to interpolation partitions (colouring the literals) in arbitrary ways.
The algorithm is implemented in SMTInterpol.

\end{abstract}

\section{Introduction}\label{sec:introduction}

Craig interpolants~\cite{DBLP:journals/jsyml/Craig57a} were originally proposed to reason about proof complexity.
In the last two decades, research reignited when interpolants proved useful for software verification, in particular for generating invariants~\cite{DBLP:conf/popl/HenzingerJMM04}.
Tree interpolants are useful for verifying programs with recursion~\cite{DBLP:conf/popl/HeizmannHP10}, and
for solving non-linear Horn-clause constraints\cite{DBLP:journals/fmsd/RummerHK15}, which can be used for thread modular reasoning~\cite{DBLP:conf/popl/GuptaPR11,DBLP:conf/popl/HoenickeMP17} and verifying array programs~\cite{DBLP:conf/sas/MonniauxG16}.
For many verification problems, reasoning about first-order quantifiers is needed. Quantifiers are, among others, needed to model unsupported theories or to express global properties of arrays~\cite{DBLP:conf/tacas/McMillan08}, for example, the sortedness~\cite{DBLP:conf/vmcai/BradleyMS06,DBLP:conf/sas/SeghirPW09}.

An interpolation problem is an unsatisfiable conjunction of several input formulas, the partitions of the interpolation problem.
An interpolant summarises the contribution of a single or multiple partitions to the unsatisfiability.
Interpolants can be computed from resolution proofs.
However, most methods require localised proofs where each literal is associated with some input partition~\cite{DBLP:conf/hvc/RolliniBS10}.
Proofs generated by SMT solvers, especially with quantifier instantiations, usually contain mixed terms and literals created during the solving process that cannot be associated with a single input formula.

In this paper, we extend our work on proof tree preserving sequence interpolation of quantified formulas \cite{DBLP:conf/smt/HenkelHS21}.
The method presented therein allows for the computation of inductive sequence interpolants from instantiation-based resolution proofs of quantified formulas in the theory of uninterpreted functions.
The key idea of this method is to perform a virtual modification of mixed terms introduced through quantifier instantiations, thus allowing to compute an inductive sequence of interpolants on a single, non-local proof tree.

We extend the interpolation algorithm to compute tree interpolants and to support arbitrary SMT theories (with the single restriction that such a theory itself must support tree interpolation for its lemmas).
We simplify the treatment of mixed terms by virtually flattening all literals independently of the partitioning.
We show that the literals can be coloured (assigned to a partition) arbitrarily, and that for every colouring, correct interpolants are produced.
The interpolants contain quantifiers for the flattening variables that bridge different partitions, and by choosing colours sensibly the number of quantifiers can be reduced.

\paragraph{Related work.}

Many practical algorithms to compute interpolants have been presented.
We focus here on proof-based methods that either work in the presence of quantifiers, or that can compute tree interpolants, or both.

Our work builds on the method presented in~\cite{DBLP:conf/dagstuhl/ChristH10} for computing interpolants from instantiation-based proofs in SMT.
It is based on \emph{purifying} quantifier instantiations by introducing variables for terms not fitting the partition, and adding defining auxiliary equalities as a new input clause in the proof.
Our method introduces these variables and equalities only virtually for computing the partial interpolants.
Thus, tree interpolants can be computed from a single proof of unsatisfiability, while in~\cite{DBLP:conf/dagstuhl/ChristH10} a purified proof is required for each partition.

There exist several methods to compute interpolants for quantified formulas
inductively from superposition-based proofs.
In~\cite{DBLP:journals/jar/BonacinaJ15}, each literal is given a \emph{label}
(similar to our colouring)
used to project the clause to the different partitions.
First, a \emph{provisional} interpolant is computed that may contain local symbols.
These symbols are replaced by quantified variables to obtain an interpolant.
In contrast to our method, the approach only works when the provisional interpolants contain at most local constants, \ie, no local functions or predicates, and the assignment of labels is not flexible as our colouring.
The method in~\cite{DBLP:conf/lpar/KovacsV17} is based on a slightly modified proof, where substitution steps are done separately.
First, a \emph{relational} interpolant is computed, which may contain local function symbols, but only shared predicates.
In logic without equality, or when the only local symbols are constants, the relational interpolant can be turned into an interpolant by quantifying over non-shared terms, respecting their dependencies.

A very different method based on summarising subproofs is presented in~\cite{DBLP:conf/cade/GleissK017}.
The proof is split into subproofs belonging to a single partition. The relevant subproofs are summarised in an \emph{intermediant} stating that their premises imply their conclusion.
If the subproofs contain only symbols of the respective partition, the resulting formula is an interpolant.
If the proof can be split in that way, the method works for any theory and proof system,
but for tree interpolation, a different proof would be required for each partitioning.

Tree interpolants can be computed by repeated binary interpolation from formulas where the children interpolants are included, as discussed in~\cite{DBLP:conf/popl/HeizmannHP10}.
In the propositional setting,
\cite{DBLP:conf/atva/GurfinkelRS13} discusses
under which conditions sets of interpolants with certain relations, such as tree interpolants, can be obtained by binary interpolation on different partitionings of the same formula.
The method is implemented in OpenSMT, but the solver, and therefore the interpolation engine, does not support quantifiers.

A general framework for computing tree interpolants for ground formulas from a single proof has been presented in~\cite{DBLP:journals/jar/ChristH16}.
It works for combinations of equality-interpolating theories and is based on projecting \emph{mixed literals} using
auxiliary variables and predicates.
Additionally, the rule for computing a resolvent's interpolant from its antecedents' interpolants is more involved.
The method cannot deal with quantifier instantiations, nor with terms mixing subterms from different partitions.
We discuss in Section~\ref{sec:equalityinterpolating} how it can be combined with the interpolation method for quantified formulas presented in this paper.

The first implementation of a tree interpolation algorithm in the presence of quantifiers and theories was in Vampire~\cite{DBLP:conf/lpar/BlancGKK13}.
It is based on repeatedly computing binary interpolants for modified interpolation problems, similar to~\cite{DBLP:conf/popl/HeizmannHP10}.
For each binary computation, the proof must be localised in order to be able to compute interpolants.
In contrast, our method computes tree interpolants in one go from a single proof that has been obtained without knowledge of the partitioning of the tree interpolation problem.
To the best of our knowledge, Vampire is the only other tool that is able to compute tree interpolants in the presence of quantifiers.

\section{Notation}\label{sec:notation}

We assume that the reader is familiar with first-order logic.
We define a \emph{theory} \T by its \emph{signature}, that contains constant, function and predicate symbols, and its set of \emph{axioms}, closed formulas that fix the meaning of those function and predicate symbols that are \emph{interpreted} by the theory.

A \emph{term} is a variable or the application of an $n$-ary function symbol to $n$ terms.
An \emph{atom} is the application of an $n$-ary predicate to $n$ terms, and a \emph{literal} is an atom or its negation.
A \emph{clause} is a disjunction of literals, and a formula is in \emph{conjunctive normal form} (CNF) if it is a conjunction of clauses.
We use $\top$ (resp.\ $\bot$) for the formula that is always true (resp.\ false).

We will demonstrate our algorithm using the theory of equality, and the theory of linear rational arithmetic.
The theory of equality establishes reflexivity, symmetry, and transitivity of the equality predicate $=$, and congruence for each \emph{uninterpreted} function symbol.
For simplicity of the presentation, uninterpreted constants are considered as $0$-ary functions, and uninterpreted predicate symbols as uninterpreted functions with Boolean return value.
The theory of linear arithmetic contains the predicates $\leq,<$, rational constants $c$, the 2-ary addition function $+$, and a family of 1-ary multiplication functions $c \cdot$, one for each rational constant $c$.
These symbols have their usual semantics and the main theory lemmas are trichotomy ($x<y \lor x=y \lor x>y$) and a variant of Farkas lemma.
For simplicity, we apply arithmetic conversions implicitly and treat $x\leq y$ and $y \geq x$ and $1\cdot x + (-1)\cdot y\leq 0$ as the same literal, and $x>y$ as its negated literal.

We denote constants by $a, b, c$, functions by $f, g, h$, variables by $v, x, y, z$, and terms by $s, t$.
We use $\ell$ for literals and $C$ for clauses, and $\phi, F, I$ for formulas.

For a term $t$, the outermost (or \emph{head}) function symbol is denoted by $\hd(t)$.
The set of all uninterpreted function symbols occurring in a formula $F$ is $\symbset(F)$ and the set of all free variables in $F$ is $\FreeVars(F)$.
The result of substituting in a formula $F$ each occurrence of a variable $x$ by a term $t$ is denoted by $F \{x \mapsto t\}$.
By $\bar{x}$ and $\bar{t}$, we denote the list of variables $x_1,\dots,x_n$ and terms $t_1,\dots t_n$, respectively. 
We use the symbol $\equiv$ to denote equivalence between formulas, and to assign a formula to a formula variable.

\section{Preliminaries}\label{sec:preliminaries}

\paragraph*{Craig interpolation.}
A binary \emph{Craig interpolant}~\cite{DBLP:journals/jsyml/Craig57a} for an 
unsatisfiable conjunction $A\land B$ is a formula $I$ that is implied by $A$, 
contradicts $B$, and contains only symbols that occur in both $A$ and 
$B$.
A generalisation are tree interpolants, which introduce several partitions in a tree-like structure.

\begin{definition}[Tree interpolation]
\label{def:treeinterpolant}
    A \emph{tree interpolation problem} $(V,E,F)$ is a labelled binary tree where $V$ is a set of nodes connected by directed edges $E\subseteq V \times V$.
    The graph is cycle-free, the root node has no outgoing edge and each other node has exactly one outgoing edge pointing towards the root node.
    The \emph{partitions} $P \subseteq V$ of the tree interpolation problem are the leaf nodes without incoming edges.  
    Each non-leaf node has exactly two incoming edges.  
    The labelling function $F$ assigns a formula to each partition $p\in P$ of the tree such that their conjunction is unsatisfiable.
    We use $\tree(v)\subseteq P$ to denote the set of leaves in the subtree of the node $v$, i.e., the set of leaves for which a path to the node $v$ exists.
    
    A \emph{tree interpolant} for the interpolation problem $(V,E,F)$ is a labelling function for all nodes with the following properties:
    \begin{enumerate}
        \item The label $I(v_r)$ of the root node $v_r$ of the tree is $\bot$.
        \item For each leaf node $p\in P$, its interpolant $I(p)$ is implied by the formula $F(p)$.
        \item For each inner node $v \in V\setminus P$, its interpolant is implied by the interpolants labelled to the two child nodes.
        \item For each node $v$, the symbols in $I(v)$ occur both inside and outside of the subtree $\tree(v)$, i.e., $\symbset(I(v)) \subseteq (\bigcup_{p \in \tree(v)} \symbset(F(p)) \cap (\bigcup_{p \not\in \tree(v)} \symbset(F(p))$.
    \end{enumerate}
\end{definition}

\paragraph{Remark.} In contrast to the earlier definition of tree interpolation~\cite{DBLP:conf/lpar/BlancGKK13,DBLP:journals/jar/ChristH16}, only the leaves of the tree are labelled here.
A tree interpolation problem with labelled inner nodes can be transformed to our formalism by adding a leaf child to each such node.
A non-binary tree can be extended to a binary tree by adding more internal nodes.
If the interpolants of the newly created nodes are ignored, the remaining interpolants are tree interpolants according to the earlier definition for tree interpolation.

Each interpolant $I(v)$ in the tree is also a binary interpolant of the formulas in the partitions $A := \tree(v)$ and $\compl{A} := P \setminus \tree(v)$.
Since the set $A$ defines $v$ uniquely, we can also use $I_A$ to denote $I(v)$.
We follow the usual convention for binary interpolation and call a symbol \emph{$A$-local} if it only occurs in partitions in~$A$, \emph{$B$-local} if it only occurs in partitions in~$\compl{A}$ and \emph{shared} if it occurs in both.
The interpolant may only contain shared symbols. 

\paragraph*{Theory combination.}
We assume Nelson--Oppen style theory combination sharing equalities without explicitly introducing auxiliary variables.  
Literals belong to one theory (except for equality) and nested terms of other theories are treated as variables.
We further assume that each theory produces their own theory lemmas and that there is a theory-specific interpolation procedure for these lemmas.
In this paper, we do not have the assumption that theories are equality-interpolating.
We introduce quantifiers in the interpolants for such theories.
However, our approach can also be combined with equality-interpolating theories and corresponding procedures to avoid quantifiers.

\paragraph*{CNF transformation and quantifiers.}
We assume that complex input formulas are transformed to CNF by introducing Boolean proxy literals and Tseitin-encoding. 
Existentially quantified variables are replaced with Skolem constants or functions (if nested under a universal quantifier) 
and conjunctions are lifted over universal quantifiers.
Thus we end up with quantified clauses of the form $\forall \bar x.\, \ell_1(\bar x) \lor \dots \lor \ell_n(\bar x)$, which
we treat as a proxy literal.
Complex subformulas under a universal quantifier are replaced by uninterpreted predicates, taking as arguments the quantified variables.
Quantified Tseitin-style axioms give the meaning for these predicates.
Instances of quantified clauses are created using instantiation lemmas of the form
$\lnot( \forall\bar x.\, \ell_1(\bar x) \lor \dots \lor \ell_n(\bar x)), \ell_1(\bar t), \dots, \ell_n(\bar t)$.
Note that the quantified formula is a proxy literal which may occur positively in input clauses.
We note that all preprocessing steps are done locally for each input formula and that auxiliary predicates and Skolem functions are fresh
function or predicate symbols.
An interpolant of the preprocessed formulas is also an interpolant of the original formulas, because the auxiliary symbols are not shared between
different input formulas and will never appear in the interpolant.

\paragraph*{Proofs.}
A \emph{resolution proof} for the unsatisfiability of a formula in CNF is a 
derivation of the empty clause $\bot$ using the resolution rule
\[\frac{C_1\lor\ell\qquad C_2\lor\lnot\ell}{C_1\lor C_2}\]
where $C_1$ and $C_2$ are clauses, and $\ell$ is a literal called the 
\emph{pivot} (literal).
A resolution proof can be represented by a tree, or more generally, if the same subproof is used more than once, by a directed acyclic graph (DAG).
In our setting, the DAG has three types of leaves:
\emph{input clauses}, \emph{theory lemmas}, i.e., clauses that are valid in the theory \T, and \emph{instantiation lemmas} of the form $\lnot(\forall\bar{x}.\phi(\bar{x}))\lor\phi(\bar{t})$.
The inner nodes are clauses obtained by resolution, and the unique root node is the empty clause $\bot$.

Binary interpolants can be computed from a resolution proof by computing so-called partial interpolants for each clause.
Each step proves a clause $C$ as a consequence from the input $A\land B$, hence it proves that $A\land B \land \lnot C$ is unsatisfiable.
If each literal in the proof is assigned to, or \emph{coloured} with, either partition $A$ or $B$, a \emph{partial interpolant} for each intermediate step is the interpolant of
$A \land \project{\lnot C}{A}$ and $B \land \project{\lnot C}{B}$, where the projection $\project{\lnot C}{A}$ extracts from the conjunction $\lnot C$ all literals that are coloured with partition $A$.
McMillan~\cite{DBLP:conf/cav/McMillan06} showed for propositional logic that partial interpolants can be computed recursively for each resolution step as the disjunction of the partial interpolants of the antecedents if the pivot is coloured as $A$, and their conjunction if it is coloured as $B$.

\section{Colouring of Terms and Literals}\label{sec:colouring}

In this section, we fix an interpolation problem $(V,E,F)$, with partitions $P\subseteq V$.
We use the following example to illustrate our interpolation algorithm.
\begin{example}[Running Example]
    Take the tree interpolation problem with nodes $V = \{123, 1, 23, 2, 3\}$ and edges $E = \{(1, 123), (23, 123), (2, 23), (3, 23)\}$ (see also Figure~\ref{fig:treeInterpolationProblem}), where the partitions $P = \{ 1, 2, 3 \}$ are labelled with $F(p) = \phi_p$ where
    \[\begin{array}{ccc}
    \phi_1 \equiv \forall x.\ g(h(x))\leq x,\quad
    & \phi_2 \equiv \forall y.\ g(y) \geq b,\quad
    & \phi_3 \equiv \forall z.\ f(g(z))\neq f(b).
    \end{array}\]
  The conjunction of the three formulas is unsatisfiable.
  Instantiating $\phi_1$ with~$b$ gives $g(h(b))\leq b$.
  Instantiating $\phi_2$ with~$h(b)$ gives $g(h(b))\geq b$.
  Together they imply $g(h(b))=b$.
  However, this contradicts $\phi_3$ instantiated with $h(b)$.
  This proof creates, among others, the new literal $g(h(b)) \leq b$.
  The term $g(h(b))$ contains function symbols that do not occur in a common partition.
  \label{ex:running}
\end{example}

We recall that by $\symbset(F(p))$ we denote the non-theory symbols occurring in the formula $F(p)$.
We also keep track of the partitions where a symbol occurs:
\begin{definition}[Partitions]
   The \emph{partitions of a function symbol} $f$ are the partitions where this symbol occurs:
   $$\partitions(f) = \{ p \in P \mid f \in \symbset(F(p))\}.$$
\end{definition}

McMillan's interpolation algorithm assumes that all symbols of a literal occur in one partition such that the literal can be coloured with that partition.
This is no longer the case in SMT, because new literals are created during the proof search, especially in the presence of instantiation lemmas.
Our solution to this problem is to split each literal into many smaller literals and 
assign each of them to a partition.
To keep the presentation simple, we flatten all (non-proxy) literals using a fresh variable for each application term.
Thus, for every term $t$ occurring in the resolution proof, we create a fresh variable $v_t$ and associate with it a set of flattening equalities.
In each literal, the top-level terms are replaced with their associated variable, and the defining equalities are conjoined.

\begin{definition}[Flattening]\label{def:flattening}
    For a term $t$, we introduce a fresh variable $v_t$, and similarly for all its subterms.
    The associated set of flattening equalities $\FlatEQ(t)$ is defined as follows:
    \begin{equation*}
        \FlatEQ(t) = \{v_{f(t_1, \ldots,t_n)} = f(v_{t_1}, \ldots, v_{t_n}) \mid \text{$f(t_1,\dots,t_n$)
        is a subterm of $t$}\}.
    \end{equation*}
    The flattened version of an equality literal $\ell \equiv t_1 = t_2$,
    is $\flatten(\ell) \equiv v_{t_1} = v_{t_2}$,
    and the associated set of flattening equalities is as follows:
    \begin{equation*}
        \FlatEQ(\ell) = \FlatEQ(t_1) \cup \FlatEQ(t_2).
    \end{equation*}
    Similarly, the flattened version of an inequality literal $\ell \equiv c_1\cdot t_1 + \dots + c_n\cdot t_n \leq c$ is
    $\flatten(\ell) \equiv c_1\cdot v_{t_1} + \dots + c_n \cdot v_{t_n} \leq c$ and the associated flattening equalities
    \begin{equation*}
        \FlatEQ(\ell) = \FlatEQ(t_1) \cup  \dots \cup \FlatEQ(t_n).
    \end{equation*}
\end{definition}
The conjunction of the equalities in $\FlatEQ(t)$ implies that $v_t=t$.
Similarly, the conjunction $\flatten(\ell) \land \bigwedge \FlatEQ(\ell)$ implies the literal $\ell$ and is equisatisfiable to $\ell$.
Proxy literals like quantified formulas are not flattened, as they will never occur in a partial interpolant.
For such a proxy literal, $\flatten(\forall x.\phi(x)) \equiv \forall x.\phi(x)$ and $\FlatEQ(\forall x.\phi(x))=\emptyset$.
\begin{example}[Flattening]\label{ex:flateq}
    Consider the literal $g(h(b))\leq b$. 
    Its flattened version is $\flatten(g(h(b))\leq b) \equiv v_{g(h(b))} \leq v_b$, and the set of flattening equalities is
    \begin{align*}
        \FlatEQ(g(h(b))\leq b) = &\quad\FlatEQ(g(h(b)))\cup \FlatEQ(b) \\
         = &\quad\{v_{g(h(b))} = g(v_{h(b)}), v_{h(b)}=h(v_b), v_b=b\}.
    \end{align*}
\end{example}

To define partial interpolants, we colour each literal $\ell$ with some partition, denoted by $\itpcolour(\ell) \in P$.
For proxy literals created during the CNF conversion, it is important to colour them with the input partition from which they were created.
The colour of other literals can be chosen arbitrarily, but a good heuristic would choose a partition where most of the outermost function symbols occur.
Each flattening equality is associated with all partitions where the corresponding function symbol occurs.
The \emph{projection} of auxiliary equations on a partition $p$, denoted by $\project{\FlatEQ(\ell)}{p}$, is defined as the conjunction
of the equalities $(v_{f(t_1,\dots,t_n)} = f(v_{t_1},\dots,v_{t_n})) \in \FlatEQ(\ell)$ where $p \in \partitions(f)$.

Finally, we define the projection of a literal $\ell$ to a partition $p$.
The \emph{projection kernel} $\minproject{\ell}{p}$ is $\flatten(\ell)$ if $p = \itpcolour(\ell)$ or $\top$ otherwise.
The \emph{projection of $\ell$ to $p$} is defined as $\project{\ell}{p} \equiv \minproject{\ell}{p} \land \project{\FlatEQ(\ell)}{p}$.
We define the projection to a set of partitions $\project{\ell}{A}$ with $A\subseteq P$ (and similarly $\minproject{\ell}{A}$) as the conjunction of all projections $\project{\ell}{p}$ with $p\in A$.
For a conjunction of literals $F\equiv \ell_1\land\dots\land\ell_n$, we define $\project{F}{p} \equiv \project{\ell_1}{p} \land \dots \land \project{\ell_n}{p}$ and similar for $\project{F}{A}$, $\minproject{F}{p}$ and $\minproject{F}{A}$.

\begin{example}[Projection of literals]\label{ex:projection}
    Consider again the literal $g(h(b))\leq b$ from our running example (Example~\ref{ex:running}), and assume that we arbitrarily assign it to partition 2, \ie, $\itpcolour(g(h(b))\leq b)=2$.
    We have $\partitions(g) = \{1,2,3\}$,  $\partitions(h) = \{1\}$ and $\partitions(b) = \{2,3\}$.
    The projections are hence
    \begin{align*}
    \project{g(h(b)\leq b}{1}& \equiv v_{g(h(b))} = g(v_{h(b)}) \land v_{h(b)} = h(v_{b})\\
    \project{g(h(b)\leq b}{2}& \equiv v_{g(h(b))} \leq v_b \land v_{g(h(b))} = g(v_{h(b)}) \land v_{b} = b\\
    \project{g(h(b)\leq b}{3}& \equiv v_{g(h(b))} = g(v_{h(b)}) \land v_{b} = b 
    \end{align*}
\end{example}

A partial interpolant of a clause $C$ is the interpolant of a slightly modified tree interpolation problem, where the projection $\project{\lnot C}{p}$ is added to each leaf node.  
Since this step adds flattening variables potentially shared between several partitions, these variables can occur in the interpolants.
The following definition accounts for the variables occurring in the projection of a clause.

\begin{definition}[Supported variable]
    We call a variable $v_t$ \emph{supported by a clause} $C$ if its corresponding  term $t$ is a subterm of a non-proxy literal $\ell$ in $C$.
\end{definition}

The partial tree interpolant of a clause $C$ may then contain a variable $v_t$ as long as it is supported by the clause $C$.

\begin{definition}[Partial tree interpolant]\label{def:partialtreeinterpolant}
    A \emph{partial tree interpolant} for a clause $C$ is a tree interpolant as defined in Definition~\ref{def:treeinterpolant} 
    for the tree interpolation problem $(V,E,F')$ where the leaves are labelled with
    $F'(p) := F(p) \land \project{\lnot C}{p}$.
    For the symbol condition, all variables supported by the clause may occur in all partial interpolants.
\end{definition}

\section{Interpolation for Quantified Formulas}
\label{sec:interpolation-quantified-formulas}

In the following, we describe how to compute tree interpolants for instantiation-based resolution proofs.
We assume that each literal has been assigned to exactly one partition of the tree interpolation problem, as described in the previous section.
Following McMillan's algorithm, we compute partial tree interpolants inductively over the proof tree. 
The leaves of the proof tree are theory lemmas, for which we use theory-specific interpolation procedures, or they are input clauses or instantiation lemmas, for which we compute partial tree interpolants as described below.
The inner nodes are obtained by resolution steps, for which we follow McMillan's algorithm to combine interpolants, and additionally treat variables that violate the symbol condition, as described later in this section.

\subsection{Interpolation Algorithm}
\label{sec:algorithm}
We start by explaining how the interpolants for leaf nodes are computed.
Our algorithm computes interpolants separately for each node $v\in V$ in the tree interpolation problem.
As mentioned in the preliminaries, we set $A=\tree(v)$ and use $I_A$ to denote the interpolant $I(v)$.

\paragraph{Input clauses.}
We assume that each input clause occurs in exactly one partition.
The partial tree interpolant for an input clause $C$ from partition $p = \itpcolour(C)$ is given by
$I_A \equiv \lnot (\minproject{\lnot C}{\compl{A}})$ if $p \in A$,
and
$I_A \equiv \minproject{\lnot C}{A}$ if $p \not\in A$.

Note that the literals can be assigned to a different partition than the clause.
Although it makes sense to assign a literal to the same partition as the input clause it occurs in, this is not possible when the literal occurs in several input clauses.
Therefore, the above formulas are not necessarily $\top$ or $\bot$.
Proxy literals always have the same colour as the input clause and will therefore never appear in the interpolant.

\paragraph{Instantiation lemmas.}
The partial tree interpolant for an instantiation lemma $C$ obtained from a quantified input clause $\forall x. \phi(x)$ from partition $\itpcolour(\forall x. \phi(x))$ is computed in the same way as for input clauses.

\paragraph{Theory lemmas.}
In principle, we only require that each theory computes a partial tree interpolant for its lemmas.
However, the interpolated formula is not exactly the original lemma because literals were flattened.
In the following, we cover transitivity, congruence, trichotomy and Farkas lemmas.

For a \emph{transitivity} lemma with the corresponding conflict
$\lnot C \equiv t_1 = t_2 \land \dots \land t_{n-1} = t_n \land t_1 \neq t_n$ we can ignore the auxiliary equations introduced by flattening the terms, as the projection kernel is also a transitivity lemma.
A partial tree interpolant is computed by summarising for each $A$ the chains of the flattened equalities (and, if applicable, the single disequality) that are assigned to a partition $p \in A$.
More precisely, let $i_1<\dots<i_m$ be the boundary indices such that $\itpcolour(t_{i_j-1}=t_{i_j})\in A$ and $\itpcolour(t_{i_j} = t_{i_j+1})\notin A$ or vice versa.
Set $i_1 = 1$ if $t_1 \neq t_n$ and $t_1=t_2$ are in different partitions and $i_m = n$  if $t_{n-1}=t_{n}$ and $t_1 \neq t_n$ are in different partitions.
If $m=0$, then all colours of the equalities are in $A$ and the interpolant is $\bot$, or they are all in $\compl{A}$ and the interpolant is $\top$.
Otherwise the interpolant summarises the equalities between the boundary indices that have a colour in $A$:
if $\itpcolour(t_1 = t_n) \notin A$, then the interpolant is $I_A \equiv v_{i_1} = v_{i_2} \land v_{i_3} = v_{i_4} \land\dots \land v_{i_{m-1}} = v_{i_{m}}$, otherwise
the interpolant is $I_A \equiv v_{i_2} = v_{i_3} \land \dots \land v_{i_{m-2}} = v_{i_{m-1}} \land v_{i_{m}} \neq v_{i_{1}}$.  Here, $v_i$ denotes the auxiliary variable introduced for $t_i$.

The flattened version of the conflict corresponding to a \emph{congruence} lemma $C \equiv f(t_1,\dots,t_n) = f(s_1,\dots,s_n) \lor t_1 \neq s_1 \lor \dots \lor t_n \neq s_n$ is
\begin{align*}
    &v_{f(t_1, \ldots, t_n)} \neq v_{f(s_1, \ldots, s_n)} \land v_{t_1} = v_{s_1} \land \ldots \land v_{t_n} = v_{s_n}\\
    &\land v_{f(t_1, \ldots, f_n)} = f(v_{t_1}, \ldots, v_{t_n}) \land v_{f(s_1, \ldots, s_n)} = f(v_{s_1}, \ldots, v_{s_n})\\
    &\land {\textstyle \bigwedge \{\ell \mid \ell \in \FlatEQ(t), t \in \{t_1, \ldots, t_n, s_1, \ldots, s_n\}\}}.
\end{align*}
Note that the formula is still a congruence conflict if we drop the last line.
Consequently, the flattening equalities for the arguments of the $f$-applications, and for their subterms, are not needed in the computation of a partial interpolant, they only establish the implication between the flattened and the original lemma.
To obtain a partial tree interpolant, we first choose an arbitrary partition $p_f\in \partitions(f)$.
The partial tree interpolant is computed as follows.
\begin{equation*}
    I_A =
    \begin{cases}
        \lnot (\minproject{\lnot C}{\compl{A}}) &\text{if $p_f\in A$}\\
        \minproject{\lnot C}{A} &\text{otherwise}
    \end{cases}
\end{equation*}

For a \emph{trichotomy} lemma $C \equiv t_1 = t_2 \lor t_1 > t_2 \lor t_1 < t_2$, both
$I_A = \minproject{\lnot C}{A}$ and $I'_A = \lnot (\minproject{\lnot C}{\compl{A}})$ are partial interpolants.  We can always choose the projection that contains at most one literal.

A partial tree interpolant for a \emph{Farkas} lemma $C \equiv \lnot \ell_1 \lor \ldots \lor \lnot \ell_n$ is computed as follows.
For simplicity we assume that all flattened literals have the form $\lnot (s_i \leq b_i)$, where $s_i$ is of the form $c_1 \cdot v_1 + \ldots + c_n \cdot v_n$ and $b_i$ is a numeric constant (a rational).
Let $k_1, \ldots, k_n$ be the Farkas coefficients, i.e., the constants $k_i > 0$ such that $\sum_{i=1}^{n} k_i \cdot s_i = 0$ and $\sum_{i=1}^{n} k_i \cdot b_i < 0$.
Then for a set of partitions~$A$, we sum up the literals $\ell_i$ with $\itpcolour(\ell_i) \in A$, multiplied by their Farkas coefficients.
We obtain $I_A = (\sum_{i, \itpcolour(\ell_i) \in A} k_i \cdot s_i) \leq (\sum_{i, \itpcolour(\ell_i) \in A} k_i \cdot b_i)$.  Variables whose coefficients sum to zero are removed from the inequality and if $A$ contains
all inequalities, which sum to the conflict $0 \leq \sum_{i=1}^{n} k_i \cdot b_i$, then $I_A = \bot$.

\begin{theorem}\label{th:pIleaf}
  The interpolants as defined in this section are valid partial tree interpolants for the respective leaf nodes.
\end{theorem}
The proof for this theorem is a straight-forward case distinction over the type of leaf node.
It is given in Appendix~\ref{sec:app_base}.

\paragraph{Resolution steps.}

In a resolution step, we obtain the partial interpolant of the resolvent using the partial interpolants of the premises.
\[
  \frac{C_1\lor\ell : I^1_A\qquad C_2\lor\lnot\ell:I^2_A}{C_1\lor C_2: I^3_A}
\]
As the first step, we follow McMillan's algorithm and combine the interpolants of the premises either with $\lor$ or with $\land$ depending on whether the pivot literal is $A$ or $B$-local.  
For tree interpolants, this is done separately for each node of the tree interpolation problem, and a literal is seen as $A$-local if its colour is one of the leaves in the subtree of the node.  

\[
  I^3_A := \begin{cases} 
  I^1_A \lor I^2_A & \text{if $\itpcolour(\ell) \in A$}\\
  I^1_A \land I^2_A & \text{if $\itpcolour(\ell) \notin A$}
  \end{cases}  
\]

The formula $I^3_A$ computed above may still contain variables supported by the antecedents that are no longer supported by the resolvent $C_1 \lor C_2$.
Each of those \emph{unsupported} variables
must either be replaced by its definition or bound by a quantifier in the partial tree interpolant.
More precisely, let $v_t$ be an unsupported variable that does not occur in the right-hand side of an equality in $\FlatEQ(t')$ for any $v_{t'} \in \FreeVars(I^3_A)$.
This variable must always exist, as there is always an outermost unsupported variable.
Let $t = f(t_1, \ldots, t_n)$. 
We replace $I^3_A$ as follows:
\begin{equation*}
    I^3_A :=
    \begin{cases}
        \exists x.\, I^3_A \{v_t \mapsto x\} & \text{if $f$ is $A$-local, \ie, } \partitions(f) \subseteq A,\\ 
        \forall x.\, I^3_A \{v_t \mapsto x\} & \text{if $f$ is $B$-local, \ie, } \partitions(f) \cap A = \emptyset,\\ 
        I^3_A \{v_t \mapsto f(v_{t_1}, \ldots, v_{t_n})\} & \text{if $f$ is shared (otherwise).}
    \end{cases}
\end{equation*}
We do this repeatedly for all variables in $\FreeVars(I^3_A)$ that are unsupported.

\begin{theorem}\label{th:pIres}
If $I_A^1$ is a partial tree interpolant of $C_1 \lor \ell$ and $I_A^2$ is a partial tree interpolant of $C_2 \lor \lnot\ell$, then $I_A^3$ as computed above, after the removal of unsupported variables, is a partial tree interpolant of $C_1 \lor C_2$.
\end{theorem}

The proof for this theorem is given in Appendix~\ref{sec:app_ind}.

\begin{example}[Resolution]\label{ex:resolutionstep}
  Take the running example and 
  suppose $\ell \equiv g(h(b)) = b$ is the pivot, $I^1_{\set{1}}  \equiv v_{g(h(b))} \leq v_b$ and $I^2_{\set{1}} \equiv \top$.
  The interpolants are combined as $I^1_{\set{1}}\land I^2_{\set{1}}$ since $\itpcolour(\ell) \not\in \set{1}$.  This results in the
  interpolant $v_{g(h(b)} \leq v_b$.
  After the resolution step, we assume that $v_{g(h(b))}, v_{h(b)}, v_g$ are no longer supported.
  The outermost variable is $v_{g(h(b))}$, which must be replaced by its definition:
  $g(v_{h(b)}) \leq v_b$.
  Now $v_{h(b)}$ is resolved and since $h$ only occurs in partition~1, an existential quantifier is used:
  $\exists y.\, g(y) \leq v_b$.
  In the final step, $v_b$ is resolved with a universal quantifier since $b$ does not occur in 1, yielding
  $\forall x\exists y.\, g(y) \leq x$.

  Note that the order of eliminating variables is important.
  If $v_b$ had been chosen in the first step despite occurring in $\FlatEQ(g(h(b))$, the resulting formula would have been $\exists y\forall x. g(y) \leq x$.
  This formula is not logical equivalent and is indeed not a valid interpolant, as it does not follow from $\forall x. g(h(x)) \leq x$.
\end{example}

\begin{theorem}\label{th:alg}
    The algorithm in this section computes valid tree interpolants from a proof of unsatisfiability.
\end{theorem}
\begin{proof}
    By induction, every node in the proof tree is labelled by a valid partial tree interpolant: Theorem~\ref{th:pIleaf} is the base case and
    Theorem~\ref{th:pIres} the inductive step.  The proof of unsatisfiability ends with the empty clause and its partial interpolant is
    a tree interpolant for the original problem.
\end{proof}

\subsection{Full Interpolation Example}
\label{sec:fullexample}

We illustrate the algorithm on our running example (Example \ref{ex:running}). Consider the tree interpolation problem given in Figure \ref{fig:treeInterpolationProblem}. 
The symbol $b$ occurs in partitions 2 and 3, $f$ in 3, g in 1, 2, and 3, and $h$ in 1.
Our goal is to compute tree interpolants $I_{\set{1}}$, $I_{\set{2}}$, and $I_{\set{3}}$ for the leaf nodes such that $\phi_1$ implies $I_{\set{1}}$, $\phi_2$ implies $I_{\set{2}}$, and $\phi_3$ implies $I_{\set{3}}$, and tree interpolant $I_{\set{2,3}}$ such that $I_{\set{2,3}}$ is implied by $I_{\set{2}}\land I_{\set{3}}$, and $I_{\set{1}} \land I_{\set{2,3}}$ implies $\bot$. 

\begin{figure}[tb]
    \centering
    \begin{tikzpicture}[node distance=0.15cm and 0.3cm, align=center]
    \node (root) [] {
      $I_{\set{1,2,3}}:\bot$};
    \node (F23) [below right= of root] {
      $I_{\set{2,3}}:\exists x.\forall y. g(y)>x$};
    
    \node (F2) [below=of F23, xshift=-2cm] {$\phi_2:\forall y.g(y) \geq b$\\$I_{\set{2}}:\forall y. g(y)\geq b$};
    \node (F3) [below=of F23, xshift=2cm] {$\phi_3:\forall z.f(g(z))\neq f(b)$\\$I_{\set{3}}:\forall y. g(y)\neq b$}; 
    \node (F1) [left=of F2] {$\phi_1: \forall x.g(h(x))\leq x$\\$I_{\set{1}}:\forall x.\exists y. g(y)\leq x$};
   
   \draw[-] (root) to (F1);
   \draw[-] (root) to (F23);
   \draw[-] (F23) to (F2);
   \draw[-] (F23) to (F3);
  \end{tikzpicture}
    \vspace{-1em}
    \caption{Tree interpolation problem from Example~\ref{ex:running} annotated by tree interpolants.}
    \label{fig:treeInterpolationProblem}
\end{figure}
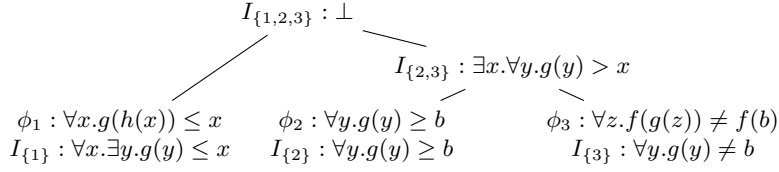

\begin{figure}[tb]
    \centering
    \begin{tikzpicture}[node distance=0.3cm and 0.22cm, align=center]
    \node (F0) [input] {$\forall x.g(h(x))\leq x$};
    \node (IF0) [interpol, right=of F0] {};
    \node (Inst0) [instance,right=of IF0] {$\lnot(\forall x.g(h(x))\leq x),~g(h(b))\leq b$};
    \node (IInst0) [interpol,right=of Inst0] {};
    \node (R0) [resolvent,below= of F0] {$g(h(b))\leq b$};   
    \node (IR0) [interpol,right=of R0] {};

    \node (T) [lemma,right=of IR0] {$g(h(b))=b, \lnot(g(h(b))\leq b),\lnot(g(h(b))\geq b)$};
    \node (IT) [interpol,right=of T] {};
    \node (R1) [resolvent,below=of R0] {$g(h(b))=b, \lnot(g(h(b))\geq b)$};
    \node (IR1) [interpol,right=of R1] {};

    \node (F1) [input,right=of IR1] {$\forall y.g(y) \geq b$};
    \node (IF1) [interpol,right=of F1] {};
    \node (Inst1) [instance,right=of IF1] {$\lnot(\forall y.g(y) \geq b),~g(h(b)) \geq b$};
    \node (IInst1) [interpol,right=of Inst1] {};
    \node (R2) [resolvent,below=of F1] {$g(h(b))\geq b$};
    \node (IR2) [interpol,right=of R2] {};

    \node (R3) [resolvent,below left=of R2, xshift=-1.3cm] {$g(h(b))=b$};
    \node (IR3) [interpol,right=of R3] {};

    \node (F2) [input,right=of IR3] {$\forall z.f(g(z))\neq f(b)$};
    \node (IF2) [interpol,right=of F2] {};
    \node (Inst2) [instance,right=of IF2] {$\lnot(\forall z.(g(z))\neq f(b)),~f(g(h(b)))\neq f(b)$};
    \node (IInst2) [interpol,right=of Inst2] {};
    \node (R4) [resolvent,below left=of Inst2] {$f(g(h(b)))\neq f(b)$};
    \node (IR4) [interpol,right=of R4] {};    

    \node (C) [lemma,right=of IR4] {$g(h(b))\neq b,~f(g(h(b))) = f(b)$};
    \node (IC) [interpol,right=of C] {};
    \node (R5) [resolvent,below=of R4] {$g(h(b))\neq b$};
    \node (IR5) [interpol,right=of R5] {};
        
    \node (bot) [resolvent,below left=of R5] {$\bot$};
    \node (Ibot) [interpol,right=of bot] {};

   \draw[->, thick] (F0) to (R0);
   \draw[->, thick] (Inst0) to (R0);
   \draw[->, thick] (R0) to (R1);
   \draw[->, thick] (T) to (R1);
       
    \draw[->, thick] (F1) to (R2);
    \draw[->, thick] (Inst1) to (R2);
    \draw[->, thick] (R1) to (R3);
    \draw[->, thick] (R2) to (R3);
   
    \draw[->, thick] (F2) to (R4);
    \draw[->, thick] (Inst2) to (R4);

    \draw[->, thick] (C) to (R5);
   \draw[->, thick] (R4) to (R5);
   
   \draw[->, thick] (R3) to (bot);
   \draw[->, thick] (R5) to (bot);

  \end{tikzpicture}
    \vspace{-1.75em}
    \caption{Resolution proof for Example~\ref{ex:running} with  \myboxIn{input clauses}, \myboxInst{\vphantom{input clauses}instantiation lemmas}, \myboxLemma{\vphantom{input clauses}theory lemmas}, and \myboxRes{\vphantom{input clauses}resolvents}.} 
    \label{fig:example-proof-tree_new}
\end{figure}
    
    Figure~\ref{fig:example-proof-tree_new} shows an instantiation-based resolution proof for the unsatisfiability of $\phi_1\land\phi_2\land\phi_3$. 
    First, we assign each literal occurring in the proof tree to exactly one partition.
    We colour each proxy literal for a quantified formula by a partition in which it occurs, \eg, $\itpcolour(\forall x. g(h(x))\leq x)=1$.
    For the other literals, we can choose arbitrary colours.
    We assign the literals $g(h(b))=b$, $g(h(b))\leq b$, and $g(h(b))\geq b$ to partition 2, and the literal $f(g(h(b)))\neq f(b)$ to partition 3.
    We then compute for each literal $\ell$ the projection onto each partition, \ie, $\project{\ell}{p_i}$. For $\ell\equiv g(h(b))\leq b$ assigned to partition 2, the projections are given in Example \ref{ex:projection}. As $g(h(b)) \geq b$ and $g(h(b)) = b$ are assigned to the same partition as $\ell$ and only differ in the comparison operator, their projections only differ in the comparison operator of the flattened version of the original literal.
    For the remaining literal $f(g(h(b))) = f(b)$, we get the projections
    \begin{equation*}
        \begin{array}{cl}
            \project{f(g(h(b))) = f(b)}{1} &\equiv v_{g(h(b))}=g(v_{h(b)}) \land v_{h(b)}=h(v_b)\\
            \project{f(g(h(b))) = f(b)}{2} &\equiv v_{g(h(b))}=g(v_{h(b)}) \land v_b=b \\
            \project{f(g(h(b))) = f(b)}{3} &\equiv v_{f(g(h(b)))}=v_{f(b)} \land v_{f(g(h(b)))}=f(v_{g(h(b))}) \land{}\\ 
            &\hphantom{{}\equiv} v_{g(h(b))}=g(v_{h(b)}) \land v_{f(b)} = f(v_b) \land v_b=b
        \end{array}
    \end{equation*}
    
    We now compute partial tree interpolants for each node in the proof tree.
    The first input clause $C \equiv \phi_1$ on the top left of the proof tree is from partition~1.
    The partial interpolant $I_{\set{1}}$ and $I_{\set{1,2,3}}$ are set to $\lnot (\minproject{\lnot C}{\compl{A}})\equiv \bot$ and $I_{\set{2}}=I_{\set{3}}=I_{\set{2,3}}$ are set to $\minproject{\lnot C}{A}\equiv \top$. 
    For the input clauses $\phi_2$ and $\phi_3$, the interpolants are computed analogously.  To summarise:

    \noindent
    \begin{tikzpicture}
      \begin{scope}[xscale=1.5]
        \node (1l) at (-1,.5) {$\phi_1:$};
        \node (1root) at (0,1) {$\bot$};
        \node (1I23) at (.5,.5) {$\top$};
        \node (1I2) at (0,0) {$\top$};
        \node (1I3) at (1,0) {$\top$}; 
        \node (1I1) at (-.75,0) {$\bot$};
        \draw[-] (1root) to (1I1);
        \draw[-] (1root) to (1I23);
        \draw[-] (1I23) to (1I2);
        \draw[-] (1I23) to (1I3);
      \end{scope}
      \begin{scope}[xscale=1.5, xshift=2.75cm]
        \node (2l) at(-1,.5) {$\phi_2:$};
        \node (2root) at (0,1) {$\bot$};
        \node (2I23) at (.5,.5) {$\bot$};
        \node (2I1) at (-.75,0) {$\top$};
        \node (2I2) at (0,0) {$\bot$};
        \node (2I3) at (1,0) {$\top$}; 
        \draw[-] (2root) to (2I1);
        \draw[-] (2root) to (2I23);
        \draw[-] (2I23) to (2I2);
        \draw[-] (2I23) to (2I3);
      \end{scope}
      \begin{scope}[xscale=1.5, xshift=5.5cm]
        \node (3l) at(-1, .5) {$\phi_3:$};
        \node (3root) at (0,1) {$\bot$};
        \node (3I23) at (.5,.5) {$\bot$};
        \node (3I1) at (-.75,0) {$\top$};
        \node (3I2) at (0,0) {$\top$};
        \node (3I3) at (1,0) {$\bot$}; 
        \draw[-] (3root) to (3I1);
        \draw[-] (3root) to (3I23);
        \draw[-] (3I23) to (3I2);
        \draw[-] (3I23) to (3I3);
      \end{scope}
    \end{tikzpicture}
    
    We now compute the partial tree interpolants for the instantiation lemma on the top right of the proof tree. Similar as for the input clauses, we set $I_{\set{1}}$ to $\lnot (\minproject{\lnot C}{\compl{A}})$,
    \ie, to $\lnot (\minproject{\lnot C}{2})\land \lnot (\minproject{\lnot C}{3})\equiv v_{g(h(b))} \leq v_b$.
    Analogously, we compute all other partial tree interpolants for the three instantiation lemmas:

    \noindent
    \begin{tikzpicture}
      \begin{scope}[xscale=3.5]
        \node (1l) at (-1.1,.8) {$\lnot\phi_1\lor g(h(b)) \leq b:$};
        \node (1root) at (0,1) {$\bot$};
        \node (1I23) at (.5,.5) {$v_{g(h(b))} > v_b$};
        \node (1I1) at (-.75,0) {$v_{g(h(b))} \leq v_b$};
        \node (1I2) at (0,0) {$v_{g(h(b))} > v_b$};
        \node (1I3) at (1,0) {$\top$}; 
        \draw[-] (1root) to (1I1);
        \draw[-] (1root) to (1I23);
        \draw[-] (1I23) to (1I2);
        \draw[-] (1I23) to (1I3);
      \end{scope}
    \end{tikzpicture}
    
      \noindent
    \begin{tikzpicture}
      \begin{scope}[xscale=1.5, yshift=-1.8cm, xshift=-1cm]
        \node (2l) at(-1.6,.7) {$\lnot \phi_2 \lor g(h(b)) \geq b:$};
        \node (2root) at (0,1) {$\bot$};
        \node (2I23) at (.5,.5) {$\bot$};
        \node (2I1) at (-.75,0) {$\top$};
        \node (2I2) at (0,0) {$\bot$};
        \node (2I3) at (1,0) {$\top$}; 
        \draw[-] (2root) to (2I1);
        \draw[-] (2root) to (2I23);
        \draw[-] (2I23) to (2I2);
        \draw[-] (2I23) to (2I3);
      \end{scope}
      \begin{scope}[xscale=1.5, yshift=-1.8cm, xshift=3.3cm]
        \node (3l) at(-1.8, .7) {$\lnot \phi_3 \lor f(g(h(b)))\neq f(b):$};
        \node (3root) at (0,1) {$\bot$};
        \node (3I23) at (.5,.5) {$\bot$};
        \node (3I2) at (0,0) {$\top$};
        \node (3I3) at (1,0) {$\bot$}; 
        \node (3I1) at (-.75,0) {$\top$};
        \draw[-] (3root) to (3I1);
        \draw[-] (3root) to (3I23);
        \draw[-] (3I23) to (3I2);
        \draw[-] (3I23) to (3I3);
      \end{scope}
    \end{tikzpicture}

    For the trichotomy lemma, the partial tree interpolants can be set to $\minproject{\lnot C}{A}$ or $\lnot(\minproject{\lnot C}{\compl{A}})$. Due to our colouring, all literals in the lemma are either in $A$ or in $\compl{A}$. 
    To get the most simple partial interpolants, we set $I_{\set{1}}$ and $I_{\set{3}}$ to $\minproject{\lnot C}{A}\equiv \top$, and $I_{\set{2}}$ and $I_{\set{2,3}}$ to $\lnot(\minproject{\lnot C}{\compl{A}})\equiv \bot$:

    \noindent
    \begin{tikzpicture}
      \begin{scope}[xscale=1.5, yshift=-2cm, xshift=-1cm]
        \node (l) at(-3.1,.7) {$g(h(b))=b \lor \lnot(g(h(b))\leq b) \lor \lnot(g(h(b))\geq b):$};
        \node (root) at (0,1) {$\bot$};
        \node (I23) at (.5,.5) {$\bot$};
        \node (I1) at (-.75,0) {$\top$};
        \node (I2) at (0,0) {$\bot$};
        \node (I3) at (1,0) {$\top$}; 
        \draw[-] (root) to (I1);
        \draw[-] (root) to (I23);
        \draw[-] (I23) to (I2);
        \draw[-] (I23) to (I3);
      \end{scope}
    \end{tikzpicture}
    
    For the congruence lemma, we have $p_f=3$.  The partial tree interpolants $I_{\set{1}}$ and $I_{\set{2}}$ are set to $\minproject{\lnot C}{A}$ as $p_f \not\in A$ for these partitions. We get $I_{\set{1}}\equiv \top$ (neither of the flattened literals in $\lnot C$ is contained in the projection kernel) and $I_{\set{2}}\equiv v_{g(h(b))}=v_b$, since we chose $2$ as the colour of this literal.
    Similarly, $I_{\set{3}}$ and $I_{\set{2,3}}$ are set to $\lnot (\minproject{\lnot C}{\compl{A}})$. 
    We get $I_{\set{3}}\equiv v_{g(h(b))}\neq v_b$ and $I_{\set{2,3}}\equiv \bot$:

    \noindent
    \begin{tikzpicture}
      \begin{scope}[xscale=1.5, yshift=-2cm, xshift=3.3cm]
        \node (l) at(-3.1, .7) {$g(h(b))\neq b \lor f(g(h(b))) = f(b):$};
        \node (root) at (0,1) {$\bot$};
        \node (I23) at (1,.5) {$\bot$};
        \node (I2) at (0,0) {$v_{g(h(b))}=v_b$};
        \node (I3) at (2,0) {$v_{g(h(b))}\neq v_b$}; 
        \node (I1) at (-1.5,0) {$\top$};
        \draw[-] (root) to (I1);
        \draw[-] (root) to (I23);
        \draw[-] (I23) to (I2);
        \draw[-] (I23) to (I3);
      \end{scope}
    \end{tikzpicture}

    Having computed the partial tree interpolants for all leaves in the proof tree, we now compute the partial tree interpolants for each resolvent.
    If the colour of the pivot literal $\ell$ is in the $A$-part, \ie, $\itpcolour(\ell) \in A$, the partial tree interpolant of the resolvent is the disjunction of the partial tree interpolants of its antecedents. Otherwise, if $\itpcolour(\ell) \in \compl{A}$, we build the conjunction of the partial tree interpolants of its antecedents. In the resolution step for the resolvent clause $C_3\equiv g(h(b)) \leq b$, the pivot literal is assigned to partition 1, \ie, $colour(\forall x. g(h(x))\leq x)=1$. 
    To obtain $I_{\set{1}}$, we hence build the disjunction of the partial interpolants of the antecedents $C_1\equiv \forall x. g(h(x))\leq x$ and $C_2\equiv \lnot(\forall x.g(h(x))\leq x)\lor~g(h(b))\leq b$, so we get  
    $I_{\set{1}}\equiv I^1_{\set{1}}\lor I^2_{\set{1}}\equiv v_{g(h(b))} \leq v_b.$
    Similarly, we obtain $I_{\set{2}}$, $I_{\set{3}}$ and $I_{\set{2,3}}$ by conjoining the respective partial interpolants.
    Since the top-left interpolant is only $\top, \bot$ and the colouring of the pivot literal ensures that we either build the conjunction with $\top$ or the disjunction with $\bot$, 
    the resulting tree interpolant of the resolvent is the same as for the top-right clause.
    The variables $v_{g(h(b))}$ and $v_b$ are both supported by $C_3$ and thus allowed to appear in the partial interpolant.
    The resolution steps of the other inner nodes are very similar in that their partial interpolants always equal
    the partial interpolant of one of their antecedents. 
    To summarise:

    \noindent
    
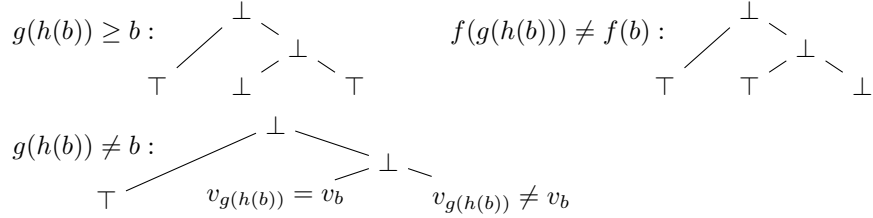
\begin{tikzpicture}
      \begin{scope}[xscale=3.5, align=left]
        \node (1l) at (-1.6,.8) {$g(h(b)) \leq b:$ \\ $g(h(b)) = b \lor \lnot(g(h(b)) \geq b):$\\ $g(h(b)) = b:$};
        \node (1root) at (0,1) {$\bot$};
        \node (1I23) at (.5,.5) {$v_{g(h(b))} > v_b$};
        \node (1I1) at (-.75,0) {$v_{g(h(b))} \leq v_b$};
        \node (1I2) at (0,0) {$v_{g(h(b))} > v_b$};
        \node (1I3) at (1,0) {$\top$}; 
        \draw[-] (1root) to (1I1);
        \draw[-] (1root) to (1I23);
        \draw[-] (1I23) to (1I2);
        \draw[-] (1I23) to (1I3);
      \end{scope}
      \end{tikzpicture}

      \noindent
    \begin{tikzpicture}
      \begin{scope}[xscale=1.5, yshift=-1.8cm, xshift=-3cm]
        \node (2l) at(-1.4,.7) {$g(h(b)) \geq b:$};
        \node (2root) at (0,1) {$\bot$};
        \node (2I23) at (.5,.5) {$\bot$};
        \node (2I1) at (-.75,0) {$\top$};
        \node (2I2) at (0,0) {$\bot$};
        \node (2I3) at (1,0) {$\top$}; 
        \draw[-] (2root) to (2I1);
        \draw[-] (2root) to (2I23);
        \draw[-] (2I23) to (2I2);
        \draw[-] (2I23) to (2I3);
      \end{scope}
      \begin{scope}[xscale=1.5, yshift=-1.8cm, xshift=1.5cm]
        \node (3l) at(-1.7, .7) {$f(g(h(b)))\neq f(b):$};
        \node (3root) at (0,1) {$\bot$};
        \node (3I23) at (.5,.5) {$\bot$};
        \node (3I2) at (0,0) {$\top$};
        \node (3I3) at (1,0) {$\bot$}; 
        \node (3I1) at (-.75,0) {$\top$};
        \draw[-] (3root) to (3I1);
        \draw[-] (3root) to (3I23);
        \draw[-] (3I23) to (3I2);
        \draw[-] (3I23) to (3I3);
      \end{scope}
      \end{tikzpicture}

      \noindent
    \begin{tikzpicture}
      \begin{scope}[xscale=1.5, yshift=-3.6cm, xshift=-2.7cm]
        \node (4l) at(-1.7, .7) {$g(h(b)) \neq b:$};
        \node (4root) at (0,1) {$\bot$};
        \node (4I23) at (1,.5) {$\bot$};
        \node (4I2) at (0,0) {$v_{g(h(b))} = v_b$};
        \node (4I3) at (2,0) {$v_{g(h(b))} \neq v_b$}; 
        \node (4I1) at (-1.5,0) {$\top$};
        \draw[-] (4root) to (4I1);
        \draw[-] (4root) to (4I23);
        \draw[-] (4I23) to (4I2);
        \draw[-] (4I23) to (4I3);
      \end{scope}
    \end{tikzpicture}

    The last resolution step is a bit more involved.
    We have already computed the tree interpolant for partition~1 in Example~\ref{ex:resolutionstep} as  $I_{\set{1}}\equiv\forall x.\exists y. g(y) \leq x$. 
    For partition~2, the disjunction $v_{g(h(b)} > v_b \lor v_{g(h(b))} = v_b$ can be simplified to $v_{g(h(b))} \geq v_b$.
    The outermost variable $v_{g(h(b))}$ is then
    replaced by $g(v_{h(b)})$, since $g$ occurs in 1 and~2.
    Then for $v_{h(b)}$ a universal quantifier is introduced, since $h$ only occurs in partition~1, resulting in $\forall y. g(y) \geq v_b$.
    Finally $v_b$ is replaced by $b$, since it occurs in both 2 and~3.  This results in $I_{\set{2}} \equiv \forall y. g(y) \geq b$.
    We omit the computation of the partial interpolant for partitions~3 and the node $23$.
    The partial tree interpolant computed in this step is the tree interpolant of the full interpolation problem:
    
    \noindent
    
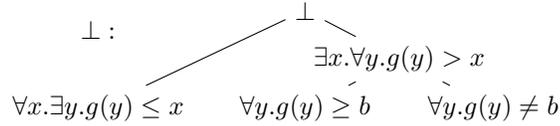
\begin{tikzpicture}
      \begin{scope}[xscale=2.5, yscale=1.3]
        \node (1l) at (-1.1,.8) {$\bot:$};
        \node (1root) at (0,1) {$\bot$};
        \node (1I23) at (.5,.5) {$\exists x.\forall y.g(y) > x$};
        \node (1I1) at (-1.1,0) {$\forall x.\exists y.g(y) \leq x$};
        \node (1I2) at (0,0) {$\forall y.g(y) \geq b$};
        \node (1I3) at (1,0) {$\forall y.g(y) \neq b$}; 
        \draw[-] (1root) to (1I1);
        \draw[-] (1root) to (1I23);
        \draw[-] (1I23) to (1I2);
        \draw[-] (1I23) to (1I3);
      \end{scope}
    \end{tikzpicture}
    
\section{Combination with Equality-Interpolating Theories}
\label{sec:equalityinterpolating}

In Sections~\ref{sec:colouring} and \ref{sec:interpolation-quantified-formulas}, we assign each literal to exactly one partition, such that we can apply McMillan's algorithm to combine partial interpolants of the antecedents to obtain a partial interpolant for the resolvent.
In the presence of equality-interpolating theories~\cite{DBLP:conf/cade/YorshM05}, we can also allow for \emph{mixed} literals where only outermost terms must be assigned to one partition.
More precisely, we can allow for equalities $t_1 = t_2$ where the left-hand side $t_1$ is in one partition and the right-hand side $t_2$ in another, or linear constraints of the form $c_1\cdot t_1 + \ldots + c_n\cdot t_n \diamond c_0$ with constants $c_i$ and $\diamond \in \{=, \leq, <, \geq, >\}$, where each $t_i$ is assigned to one partition.
Such literals can be treated by applying proof tree preserving tree interpolation~\cite{DBLP:journals/jar/ChristH16}.

A mixed literal $\ell \equiv t_1=t_2$ is coloured with two colours $p_1$ and $p_2$, so that each colour can be chosen to contain the outermost symbols of $t_1$ and $t_2$, respectively.
The projections are $\minproject{\ell}{p_1} \equiv v_{t_1} = v_{\ell}$, $\minproject{\ell}{p_2} \equiv v_{\ell} = v_{t_2}$  and for the
negated literal $\minproject{\lnot\ell}{p_1} \equiv EQ_1(v_{\ell}, v_{t_1})$ and $\minproject{\lnot\ell}{p_2} \equiv EQ_2(v_{\ell}, v_{t_2})$, where $v_{\ell}$ is a fresh variable and $EQ_1,EQ_2$ are shared uninterpreted predicates with $\forall x,y.\lnot(EQ_1(x,y) \land EQ_2(x,y))$ that are only used for the interpolation algorithm.
The resolution step uses syntactical rules to remove the auxiliary variable $v_\ell$ without introducing quantifiers, similar to what destructive equality reasoning would do.
This method produces quantifier-free interpolants if the input formulas were quantifier-free.

\section{Implementation in SMTInterpol}
\label{sec:implementation}

We implemented the algorithm in SMTInterpol~\cite{DBLP:conf/spin/ChristHN12} with a few alterations.  First, we used the combination with equality-interpolating theories described in the previous section.
Second, we do not apply flattening explicitly.
Instead of using an auxiliary variable, the interpolation algorithms for the lemmas includes the corresponding term directly.  This may result in an interpolant where the interpolant has symbols that are not allowed, because the auxiliary variable was shared but its corresponding function symbol is local to one partition.  
Only in that case, we introduce the fresh variables for these subterms and replace the offending subterm in the interpolant with its variable.   This creates the same interpolants as our presented algorithm, because the latter replaces each variable that stands for a shared function symbol by its definition in the end.

SMTInterpol also supports literals that are shared.
If this is done naïvely, the computed interpolants may violate the tree inductivity property (third property in Definition~\ref{def:treeinterpolant}).
We solve this by using tree interpolation procedures for each lemma that treat each literal as occurring in one designated partition (minimizing the number of alternating chains in transitivity lemmas).
Our implementation colours input literals with all partitions it occurs in.
For new terms created in the proof, the colour that matches the most outermost function symbols is chosen.
If the term uses only symbols from one partition, then it is coloured with that partition.
Equalities and inequalities between terms of different partitioned are handled with the equality-interpolating procedure to avoid introducing quantifiers when it is not necessary.

\section{Conclusion}
\label{sec:conclusion}
We presented a tree interpolation algorithm for SMT formulas with quantifiers.
The key idea is to virtually flatten each conflict corresponding to a clause in the resolution proof, such that the literals in the flattened version are non-mixed and can be assigned to the different partitions.
The colouring of the original literals can even be chosen arbitrarily.
Depending on the assigned colours, partial interpolants may contain flattening variables that bridge different partitions, which eventually must be bound by quantifiers. 

Our algorithm computes tree interpolants from a single, non-local proof of unsatisfiability obtained independently of the partitioning of the interpolation problem.
It supports quantifiers and arbitrary SMT theories, given that the theory itself supports tree interpolation for its lemmas, and we provided the algorithms for the theory of equality and the theory of linear rational arithmetic.

Correctness proofs for our algorithm are available in Appendix~\ref{sec:appendix}.
The algorithm is implemented in the open-source SMT solver SMTInterpol.

\newpage
\begin{appendix}
    \section{Proofs}\label{sec:appendix}
This appendix contains the complete proofs to establish correctness of our interpolation algorithm, as stated in Theorems~\ref{th:pIleaf}, \ref{th:pIres}, and \ref{th:alg}.

Let $(V,E,F)$ be the tree interpolation problem defining partitions $P$.
We prove the following properties to show that a formula is a partial interpolant for a clause $C$:
\begin{enumerate}
    \item For the root node of the tree interpolation problem, associated with the complete set of partitions $P$, we have
    \begin{equation}
        I_{P} \equiv \bot\tag{root condition}\label{eq:root-cond}
    \end{equation}
    \item For each partition $p \in P$, the following implication is valid:
    \begin{equation}
        F(p) \land \project{\lnot C}{p} \rightarrow I_{\{p\}}\tag{leaf-ind}\label{eq:base-ind}
    \end{equation}
    \item For each two disjoint subtrees defining sets of partitions $A_1$, $A_2$, the following implication is valid:
    \begin{equation}
        I_{A_1} \land I_{A_2} \rightarrow I_{A_1 \dot\cup A_2}\tag{tree-ind}\label{eq:tree-ind}
    \end{equation}
    \item For each set of partitions $A$, the formula $I_A$ contains only symbols that occur in both $A$ and $\compl{A}$ and variables supported by the current clause $C$, \ie,
    \begin{equation}
        \symbset(I_A) \subseteq \biggl( \Bigl( \bigcup_{p \in A} \symbset(F(p)) \Bigr) \cap \Bigl( \bigcup_{p \in \compl{A}} \symbset(F(p)) \Bigr) \biggr) \cup \Supported(C) \tag{symbol condition}\label{eq:symb-cond}
    \end{equation}
    where $\Supported(C)$ denotes the set of variables supported by $C$.
\end{enumerate}
Note that the property~\eqref{eq:tree-ind} is stronger than the corresponding property in the definition of a partial tree interpolant, Definition~\ref{def:partialtreeinterpolant}, because it holds for arbitrary disjoint sets $A_1$ and $A_2$.


\subsection{Proof for Theorem \ref{th:pIleaf}}\label{sec:app_base}
Theorem~\ref{th:pIleaf} states that our algorithm computes partial tree interpolants for the leaf nodes.
We give separate proofs for input clauses, instantiation lemmas, and theory lemmas.

\subsection*{Input clauses}

Let $C$ be an input clause.
The symbol condition holds trivially for any $I_A$ as the projection kernel (for any set of partitions $A$) of a clause contains only variables supported by the clause.
The interpolant of the root of the tree interpolation problem is $I_P \equiv \lnot (\minproject{\lnot C}{\compl{P}}) \equiv \lnot (\minproject{\lnot C}{\emptyset}) \equiv \bot$.

\paragraph*{Leaf inductivity.} Next, we show that \eqref{eq:base-ind} holds.
Let $p \in P$ be a partition of the tree interpolation problem.
We have to consider two cases: either $C$ is from $p$, i.e., $\itpcolour(C)=p$, or $C$ is from a different partition $q \in P \setminus \set{p}$.
\begin{enumerate}
    \item Case $\itpcolour(C) = p$:
    Our algorithm computes $I_{\set{p}} \equiv \lnot (\minproject{\lnot C}{(P \setminus \set{p})})$, thus we have to show that
    $F(p)\land \project{\lnot C}{p}\rightarrow \lnot (\minproject{\lnot C}{(P \setminus \set{p})})$
    is valid, or, equivalently, that the following formula is unsatisfiable.
    \begin{equation}
        F(p)\land \project{\lnot C}{p} \land \minproject{\lnot C}{(P \setminus \set{p})}
        \label{eq:input_base-ind_A}
    \end{equation}
    As $C$ is from partition $p$, it is one of the conjuncts of $F(p)$, and $F(p) \equiv R \land C$ for some conjunction $R$.
    Furthermore, as all function symbols in $C$ occur in $F(p)$, the projection $\project{\lnot C}{p}$ contains all flattening equalities of the literals in $C$.
    Therefore, the following equivalence holds.
    \begin{equation*}
        \project{\lnot C}{p}\land\minproject{\lnot C}{(P\setminus \{p\})} \equiv \project{\lnot C}{p}\land\project{\lnot C}{(P\setminus \{p\})}
    \end{equation*}
    Furthermore, $\project{\lnot C}{p} \land \project{\lnot C}{(P\setminus \{p\})}$ implies $\lnot C$.
    
    It follows that \eqref{eq:input_base-ind_A} implies $R \land C \land \lnot C$,
    and this formula is clearly unsatisfiable.
    
    \item Case $\itpcolour(C)\in P \setminus \set{p}$:
    Our algorithm computes $I_{\set{p}} \equiv \minproject{\lnot C}{p}$,
    thus we have to show validity of the following implication.
    \begin{align*}
        F(p)\land \project{\lnot C}{p}\rightarrow \minproject{\lnot C}{p}
    \end{align*}
    As the projection kernel is implied by the projection, this holds trivially.
\end{enumerate}

\paragraph*{Tree inductivity.} It remains to show that \eqref{eq:tree-ind} holds.
Let $A_1$ and $A_2$ be sets of partitions defined by two disjoint subtrees of the tree interpolation problem $(V,E,F)$.
We have to consider three cases: either $\itpcolour(C) \in A_1$ or $\itpcolour(C) \in A_2$, or $\itpcolour(C)\in \compl{(A_1 \dot\cup A_2)}$.

\begin{enumerate}
    \item Case $\itpcolour(C)\in A_1$:
    Our algorithm computes the following formulas.
        \begin{align*}
            I_{A_1} \equiv \lnot(\minproject{\lnot C}{\compl{A_1}}),
            \quad I_{A_2} \equiv \minproject{\lnot C}{A_2},
            \quad I_{A_1 \dot\cup A_2} \equiv \lnot(\minproject{\lnot C}{\compl{(A_1 \dot\cup A_2)}})
        \end{align*}
        We have to show validity of the following implication.
        \[\lnot(\minproject{\lnot C}{\compl{A_1}}) \land \minproject{\lnot C}{A_2} \rightarrow \lnot(\minproject{\lnot C}{\compl{(A_1 \dot\cup A_2)}})\]
        Rearranging the formula and using $\compl{A_1}=A_2\dot\cup \compl{(A_1\dot\cup A_2)}$ we get
        \begin{align*}
            & \minproject{\lnot C}{A_2} \land \minproject{\lnot C}{\compl{(A_1 \dot\cup A_2)}} \rightarrow \minproject{\lnot C}{\compl{A_1}}\\
            \equiv \quad & \minproject{\lnot C}{\compl{A_1}} \rightarrow \minproject{\lnot C}{\compl{A_1}}
        \end{align*}
        which trivially holds.
    \item Case $\itpcolour(C)\in A_2$ is symmetrical to the first case.
    \item Case $\itpcolour(C)\in \compl{(A_1 \dot\cup A_2)}$:
    Our algorithm computes the following formulas.
    \begin{align*}
        I_{A_1} \equiv \minproject{\lnot C}{A_1},
        \quad I_{A_2} \equiv \minproject{\lnot C}{A_2},
        \quad I_{A_1 \dot\cup A_2} \equiv \minproject{\lnot C}{(A_1 \dot\cup A_2)}
    \end{align*}
    As the conjunction of the first two formulas is equal to the third (modulo reordering of literals), the implication $ I_{A_1} \land I_{A_2} \rightarrow I_{A_1 \dot\cup A_2}$ is valid.
    \ignore{
        \begin{align*}
            I_{A_1} \equiv I_{A_2} \equiv I_{A_1 \dot\cup A_2} \equiv \minproject{\lnot C}{(A_1 \dot\cup A_2)}
        \end{align*}
        Thus, $ I_{A_1} \land I_{A_2} \rightarrow I_{A_1 \dot\cup A_2}$ trivially holds.
    }
\end{enumerate}

\subsection*{Instantiation lemmas}
Let $C$ be an instantiation lemma, i.e., it is of the form $\lnot(\forall\bar{x}.C') \lor C'(\bar{t})$.
The properties~\eqref{eq:symb-cond} and \eqref{eq:root-cond} hold for the same reasons as for input clauses.

\paragraph*{Leaf inductivity.} Next, we show that \eqref{eq:base-ind} holds.
\begin{enumerate}
    \item Case $\itpcolour(\forall\bar{x}.C')=p:$
    Our algorithm computes $I_{\set{p}} \equiv \lnot(\minproject{\lnot C}{(P \setminus \set{p})})$.
    \ignore{
        \begin{align*}
            F(p)\land \project{\lnot C}{\set{p}}\rightarrow \lnot (\minproject{\lnot C}{\compl{A}}).
        \end{align*}
    }
    Analogously to the first case for input clauses, we prove that the following formula is unsatisfiable.
    \begin{align*}
        F(p)\land \project{\lnot C}{p} \land \minproject{\lnot C}{(P\setminus\{p\})} 
    \end{align*}
    For all literals in $\lnot C$, their flattened version is contained either in $\project{\lnot C}{p}$ or in $\minproject{\lnot C}{(P\setminus\{p\})}$.
    The projection onto partition $p$ contains at least the flattening equalities for all symbols that occur in the quantified literal. 
    For an instantiated term $t$ with $p\notin\partitions(hd(t))$, the flattening equality defining $v_t$ is not contained in any of the two projections above.
    \ignore{
        However, the quantified literal implies $C[\bar{x}/\bar{v_t}]$. \TODO{is the following understandable?} By transitivity, we thus get a new valid conflict on the left side of the implication. The implication is hence satisfied.
    }
    Such a term can only have been created by instantiating a variable $x$ with $t$.
    The quantified formula, which is part of $F(p)$, also implies $C'(\bar{v_t})$.
    Using the flattening equalities for the terms $t'$ with $p \in \partitions(hd(t'))$ and transitivity, we get a contradiction with $\project{\lnot C}{p} \land \minproject{\lnot C}{(P\setminus\{p\})}$.

    \item Case $\itpcolour(\forall\bar{x}.C')\neq p$: Analogous to second case for input clauses.
\end{enumerate}

\paragraph*{Tree inductivity.} The proof that \eqref{eq:tree-ind} holds for the partial interpolants our algorithm computes for instantiation lemmas
is analogous to the proof for input clauses.

\subsection*{Congruence lemmas}

Let $C$ be a congruence lemma.
Again, the properties~\eqref{eq:symb-cond} and \eqref{eq:root-cond} hold for the same reasons as for input clauses.

\paragraph*{Leaf inductivity.} We show that \eqref{eq:base-ind} holds.
\begin{enumerate}
    \item Case $p = p_f$:
    Our algorithm computes $I_{\set{p}} \equiv \lnot (\minproject{\lnot C}{P \setminus \set{p}})$.
    Similarly to the first case for input clauses, we show that the following formula is unsatisfiable.
    \ignore{
        \begin{align*}
            F(p)\land \project{\lnot C}{\set{p}}\rightarrow \lnot (\minproject{\lnot C}{\compl{\set{p}}}).
        \end{align*}
        Rearranging the equation and using $\compl{A}=P\setminus\{p\}$ we get 
    }
    \begin{align*}
        F(p)\land \project{\lnot C}{p} \land \minproject{\lnot C}{(P\setminus\{p\})}
    \end{align*}
    We use that $\project{\lnot C}{p}$ contains at least
    $\minproject{\lnot C}{p}$ and the auxiliary equalities $v_{ft}=f(t_1,\dots,t_n)$ 
    and $v_{fs}=f(s_1,\dots,s_n)$, because $p=p_f \in \partitions(f)$.
    The kernel of the projection of a congruence conflict $C$ onto all partitions together with the flattening equalities for $v_{ft}$ and $v_{fs}$ (but without the flattening equalities for $v_{s_i}$ and $v_{t_i}$),
    is itself a congruence conflict, \ie, the following implication is valid.
    \begin{align*}
         v_{ft}=f(t_1,\dots,t_n) \land v_{fs}=f(s_1,\dots,s_n) \land
         \minproject{\lnot C}{P} \rightarrow \bot
    \end{align*}
    Therefore, the above formula is unsatisfiable.

    \item Case $p \neq p_f$: Analogous to the second case of input clauses.
\end{enumerate}

\paragraph*{Tree inductivity.} Now, we show that \eqref{eq:tree-ind} holds..
\begin{enumerate}
    \item Case $p_f \in A_1$: Analogous to first case of \eqref{eq:tree-ind} for input clauses.
    \item Case $p_f \in A_2$: Analogous to second case of \eqref{eq:tree-ind} for input clauses.
    \item Case $p_f \in \compl{(A_1\dot\cup A_2)}$: Analogous to third case of \eqref{eq:tree-ind} for input clauses.
\end{enumerate}

\subsection*{Transitivity lemmas}
Let $C$ be a transitivity lemma $C$ corresponding to the conflict 
$t_1=t_2\land \dots \land t_{n-1}=t_n\land t_1 \neq t_n$.
We define $i_1,\dots,i_m$ and $v_1,\dots,v_n$ as in Section~\ref{sec:algorithm}.

The symbol condition holds, as the variables $v_1,\dots,v_n$ used to summarise chains of (dis-)equalities assigned to $A$ are all supported by the clause.
For the root node of the tree interpolation problem, we get $I_P \equiv \bot$ as all chains are assigned to $P$ and the flattened literals form a transitivity conflict themselves.

\paragraph*{Leaf inductivity.} Next, we show that \eqref{eq:base-ind} holds for partial interpolants of transitivity lemmas.
The case where $m = 0$ holds is trivial: $\top$ is always implied,
and if $(\minproject{\lnot C}{p})$ contains all literals, it is a negated transitivity lemma and implies $\bot$, as for the root node.
For $m > 0$ we have to consider two cases.
\begin{enumerate}
    \item Case $\itpcolour(t_1 = t_n) \in P \setminus \set{p}$:
    We have to show
    \begin{align*}
        F(p) \land \minproject{\lnot C}{p} \rightarrow 
    v_{i_1} = v_{i_2} \land v_{i_3} = v_{i_4} \land\dots \land v_{i_{m-1}} = v_{i_{m}}.
    \end{align*}
    The equality $v_{i_1}=v_{i_2}$ is implied because $\minproject{\lnot C}{p}$ contains the equalities $v_{i_1}=v_{i_1+1},\ldots, v_{i_2-1}=v_{i_2}$ by definition of $i_1,i_2$.
    This holds analogously for $v_{i_3} = v_{i_4}$ and so on.

    \item Case $\itpcolour(t_1 = t_n) = p$:
    The proof is analogous to the first case.
    The last literal
    $v_{i_m} \neq v_{i_1}$ in the interpolant $I_{\set{p}}$ is implied because
    $\minproject{\lnot C}{p}$ contains
    $v_{i_m}=v_{i_m+1},\dots,v_{n-1}=v_n, v_1 \neq v_n, v_1=v_2, \dots,v_{i_1-1} = v_{i_1}$.
\end{enumerate}

\paragraph*{Tree inductivity.} Finally, we show that \eqref{eq:tree-ind} holds.
First note that since $A_1$ and $A_2$ are disjoint, the equality chains that are collected for $I_{A_1}$ and $I_{A_2}$ are disjoint.  
We look at the interpolant for $I_{A_1\dot\cup A_2}$ and define $i_1,\dots,i_m$ for the set $A_1\dot\cup A_2$.
If $m=0$ and $I_{A_1\dot\cup A_2}=\bot$, then all literals have a colour either in $A_1$ or $A_2$.
The sequences $i'_1<\dots<i'_{m'}$ for the interpolants $A_1$ and $A_2$ are identical 
because these are the positions
where one of the literals $v_{i'_j-1}=v_{i'_j}$ and $v_{i'_j} = v_{i'_j+1}$ is in $A_1$ and the 
other in $A_2$.
By definition (after reordering the literals) the following holds.
\begin{equation*}
I_{A_1} \land I_{A_2} \equiv 
v_{i'_1}=v_{i'_2} \land v_{i'_2}=v_{i'_3}\land \dots \land v_{i'_{m'-1}}=v_{i'_{m'}} \land v_{i'_{m'}} \neq v_{i'_1}
\end{equation*}
Since this is itself a negated transitivity lemma, it implies $I_{A_1\dot\cup A_2} \equiv \bot$.

For $m>0$ we consider the following two cases.
\begin{enumerate}
    \item Case $\itpcolour(t_1=t_n) \notin (A_1 \dot\cup A_2)$:
    We show that $v_{i_1}=v_{i_2}$ is implied by $I_{A_1}$ and $I_{A_2}$.
    The remaining conjuncts are implied for the same reason.
    We first note that $i_1$ is a boundary indices in either $I_{A_1}$ or $I_{A_2}$ and so is $i_2$.
    Furthermore, each index $j$ with $i_1 < j < i_2$ is either a boundary index in both $A_1$ and $A_2$ (where the previous literal was in one partition and the next literal in the other) or in none of them.
    Let $i_1 < j_1 < \dots < j_{m'} < i_2$ be all boundary indices in $A_1$, $A_2$ between $i_1$ and $i_2$.
    Then $I_{A_1}$ and $I_{A_2}$ together contain the equalities
    $v_{i_1} = v_{j_1} \land \dots \land v_{j_{m'}} = v_{i_2}$   (or just $v_{i_1} = v_{i_2}$ if $m'=0$).
    These equalities imply $v_{i_1}=v_{i_2}$.
    \item Case $\itpcolour(t_1=t_n) \in (A_1 \dot\cup A_2)$:
    The proof is analogous to the first case.
    For the last literal $v_{i_m} \neq v_{i_1}$ one again collects exactly one disequality similar to \eqref{eq:base-ind}.
\end{enumerate}

\subsection*{Trichotomy lemmas}
We first remark that for a trichotomy lemma $C$ the formulas $\minproject{\lnot C}{A}$ and $\lnot (\minproject{\lnot C}{\compl{A}})$ are equivalent since $(\minproject{\lnot C}{A}) \land (\minproject{\lnot C}{\compl A})\leftrightarrow \bot$.  
So it suffices to show the properties for $\minproject{\lnot C}{A}$.

The symbol condition holds for any $I_A$ for the same reason as for input clauses.
The interpolant of the root node of the tree interpolation problem is $I_P\equiv \lnot(\minproject{\lnot C}{\compl{P}})\equiv\bot$ (the negation of the empty conjunction).

\paragraph*{Leaf inductivity.}
We show that \eqref{eq:base-ind} holds for partial interpolants of trichotomy lemmas, \ie, we show 
    \begin{align*}
        F(p)\land \project{\lnot C}{p}\rightarrow\minproject{\lnot C}{p}.
    \end{align*}
The proof is analogous to the second case of \eqref{eq:base-ind} for input clauses.

\paragraph*{Tree inductivity.}
Now, we show that \eqref{eq:tree-ind} holds. We have 
\[I_{A_1}=\minproject{\lnot C}{A_1},\qquad I_{A_2}=\minproject{\lnot C}{A_2},\qquad I_{A_1\dot\cup A_2}=\minproject{\lnot C}{(A_1\dot\cup A_2)}.\]
The implication
\[\minproject{\lnot C}{A_1} \land \minproject{\lnot C}{A_2} \rightarrow \minproject{\lnot C}{(A_1\dot\cup A_2)}\]
is trivially satisfied since its left side can be simplified to $\minproject{\lnot C}{(A_1 \dot\cup A_2)}$.

\subsection*{Farkas lemmas}
Let $C:\equiv \lnot l_1 \lor\dots\lor \lnot l_n$ be a Farkas lemma and let all flattened literals $l_i$ be of the form $s_i\leq b_i$ as described in Section~\ref{sec:algorithm}.
The symbol condition holds, as all variables introduced during flattening are supported by the clause.
The interpolant of the root node of the interpolation problem is $I_P\equiv \bot$ as all inequalities are contained in $P$.

\paragraph*{Leaf inductivity.}
We show that \eqref{eq:base-ind} holds for partial interpolants of Farkas lemmas. 
First, note that the projection $\project{\lnot C}{p}$ contains the conjunction of all flattened literals $l_i$ of the form $s_i\leq b_i$ that are assigned to partition $p$, \ie, for which $\itpcolour(l_i)=p$ holds. The projection hence implies $\bigwedge_{i, \itpcolour(\ell_i)=p} s_i\leq b_i$. 

We make a case distinction on whether all literals in $C$ are assigned to partition $p$ or not.
\begin{enumerate}
    \item Case ``all literals assigned to $p$'': Our algorithm sets $I_{\set{p}}$ to $\bot$. Thus, we have to show the validity of the implication $F(p)\land\project{\lnot C}{p}\rightarrow\bot$ or, equivalently, the unsatisfiability of the following formula.
    \begin{equation}
        F(p)\land\project{\lnot C}{p}\label{eq:leaf-ind-Farkas-bot}
    \end{equation}
    The projection implies $\bigwedge_{i=1}^{n} s_i\leq b_i$, which is itself a Farkas conflict with the same coefficients as the original one.
    Thus, \eqref{eq:leaf-ind-Farkas-bot} is unsatisfiable.

    \item Case ``otherwise'': We show that \eqref{eq:base-ind} holds, which is
    \begin{equation}
        F(p)\land\project{\lnot C}{p} \rightarrow \left( (\sum_{i, \itpcolour(\ell_i)=p} k_i \cdot s_i) \leq (\sum_{i, \itpcolour(\ell_i)=p} k_i \cdot b_i)\right)\label{eq:leaf-ind-farkas}
    \end{equation}
    The projection $\project{\lnot C}{p}$ contains $s_i \leq b_i$  for all literals with $\itpcolour(\ell_i) = p$. 
    Also for the Farkas coefficients $k_i\geq 0$ holds.  Thus the implication holds.
    Note that removing the variables whose coefficients sum up to zero does not affect the truth value of the interpolant.
\end{enumerate}

\paragraph*{Tree inductivity.}
Now, we show that \eqref{eq:tree-ind} holds. If $A_1\dot\cup A_2$ does not contain all inequalities, we have
    \begin{align*}
        I_{A_1}=&(\sum_{i, \itpcolour(\ell_i)\in A_1} k_i \cdot s_i) \leq (\sum_{i, \itpcolour(\ell_i)\in A_1} k_i \cdot b_i)\\
        I_{A_2}=&(\sum_{i, \itpcolour(\ell_i)\in A_2} k_i \cdot s_i) \leq (\sum_{i, \itpcolour(\ell_i)\in A_2} k_i \cdot b_i)\\
        I_{A_1\dot\cup A_2}=&(\sum_{i, \itpcolour(\ell_i)\in A_1\dot\cup A_2} k_i \cdot s_i) \leq (\sum_{i, \itpcolour(\ell_i)\in A_1\dot\cup A_2} k_i \cdot b_i)
    \end{align*}
    The implication $I_{A_1}\land I_{A_2}\rightarrow I_{A_1\dot\cup A_2}$ is satisfied since $I_{A_1\dot\cup A_2}$ is just the
    sum of the two inequalities $I_{A_1}$ and $I_{A_2}$, because the literals in $A_1$ and $A_2$ are disjoint and $A_1\dot\cup A_2$ is their union.

    If $A_1\dot\cup A_2$ contains all inequalities, then the inequalities $I_{A_1}$ and $I_{A_2}$ sum up to
    $0  = (\sum_i k_i \cdot s_i) \leq \sum_i k_i \cdot b_i$, which implies $I_{A_1\dot\cup A_2}=\bot$ because $\sum_i k_i\cdot b_i < 0$.

\subsection{Proof for Theorem \ref{th:pIres}}\label{sec:app_ind}
Theorem~\ref{th:pIres} states that our algorithm computes partial tree interpolants for the resolvents of a resolution step, given partial tree interpolants for the antecedents. 

By the induction hypothesis, the partial interpolant of the root node of the tree interpolation problem is $I_P\equiv\bot$ for the antecedents. In the resolution step, the interpolants are combined by disjunction, thus preserving $I_P\equiv\bot$.

In general, the partial interpolants of the antecedents only contain shared symbols or supported variables. The combined partial interpolant may then contain variables that are no longer supported. As each unsupported variable is either replaced by its definition or bound by a quantifier in the last step of our algorithm, the symbol condition is established.

\subsection*{Leaf Inductivity}  
We split the proof into two parts. First, we show that 
the property holds after the McMillan step of our algorithm,
provided we still assume that all auxiliary equalities introduced by $\ell$ are true.
Then, we show that after quantifier introduction and variable replacement, \eqref{eq:base-ind} holds without assuming additional equalities.

\paragraph*{McMillan step.}
Let $p \in P$ be a partition of the tree interpolation problem.
Further, let $C_3=C_1\lor C_2$ be the resolvent for the resolution of the antecedent clauses $C_1\lor \ell$ and $C_2\lor\lnot\ell$ and $I^1$ and $I^2$ the partial tree interpolants of the antecedents. 

As $I^1$ and $I^2$ are valid partial tree interpolants, we know that
\begin{align}
    F(p) \land \project{\lnot(C_1\lor\ell)}{p}&\rightarrow I^1_{\set{p}}\label{eq:leaf-ind-C1}\\
    F(p) \land \project{\lnot(C_2\lor\lnot\ell)}{p}&\rightarrow I^2_{\set{p}} \label{eq:leaf-ind-C2}
\end{align}
hold. 
We show that the following implication holds after the McMillan step of our algorithm.
\begin{align}
F(p) \land \project{\lnot C_3}{p} \land \project{\FlatEQ(\ell)}{p}\rightarrow I^3_{\set{p}}\label{eq:leaf-ind-with-auxeq}
\end{align}
Here, we still have the auxiliary equalities introduced by the literal $\ell$ on the left-hand side.
We have to consider two cases depending on the pivot literal $\ell$ of the resolution step: either $\ell$ is assigned to $p$, \ie, $\itpcolour(\ell)=p$, or $\ell$ is assigned to a different partition $q \in P \setminus \set{p}$.

\begin{enumerate}
    \item Case $\itpcolour(\ell)=p:$
    First we note that \eqref{eq:leaf-ind-C1}~and~\eqref{eq:leaf-ind-C2} can be written as
    \begin{align}
      F(p) \land \project{\lnot C_1}{p}\land \lnot\flatten(\ell) \land \project{\FlatEQ(\ell)}{p} &\rightarrow I^1_{\set{p}} \label{eq:leaf-ind-C1-Alocal}\\
      F(p) \land \project{\lnot C_2}{p}\land \flatten(\ell) \land \project{\FlatEQ(\ell)}{p}&\rightarrow I^2_{\set{p}}.  \label{eq:leaf-ind-C2-Alocal}
    \end{align}
    To prove \eqref{eq:leaf-ind-with-auxeq}, we assume that the following holds.    \[F(p)\land \project{\lnot C_3}{p} \land \project{\FlatEQ(\ell)}{p}\]
    Using $C_3=C_1\lor C_2$, we can split up the projection, hence
    $\project{\lnot C_1}{p}$ and ${\project{\lnot C_2}{p}}$ both hold.
    If $\flatten(\ell)$ does not hold, then $I^1_{\set{p}}$ holds by \eqref{eq:leaf-ind-C1-Alocal}.
    Otherwise, $I^2_{\set{p}}$ holds by \eqref{eq:leaf-ind-C2-Alocal}.
    Hence in both cases $I^1_{\set{p}} \lor I^2_{\set{p}}$ holds and we have shown
    \[F(p)\land \project{\lnot C_3}{p} \land \project{\FlatEQ(\ell)}{p}\rightarrow I^1_{\set{p}}\lor I^2_{\set{p}}.\]

    \item Case $\itpcolour(\ell)=P\setminus \set{p}:$
    As the pivot literal $\ell$ is assigned to a different partition than $p$, 
    we can simplify \eqref{eq:leaf-ind-C1}~and~\eqref{eq:leaf-ind-C2} to
    \begin{align}
      F(p) \land \project{\lnot C_1}{p}\land \project{\FlatEQ(\ell)}{p}& \rightarrow I^1_{\set{p}} \label{eq:leaf-ind-C1-Blocal}\\
      F(p) \land \project{\lnot C_2}{p}\land \project{\FlatEQ(\ell)}{p}&\rightarrow I^2_{\set{p}}.  \label{eq:leaf-ind-C2-Blocal}
    \end{align}
    Again, we assume that the following holds.
    \[F(p)\land \project{\lnot C_3}{p} \land \project{\FlatEQ(\ell)}{p}\]
    And as mentioned above this means that
    $\project{\lnot C_1}{p}$ and $\project{\lnot C_2}{p}$ both hold.
    Now, $I^1_{\set{p}} \land I^2_{\set{p}}$ holds by \eqref{eq:leaf-ind-C1-Blocal} and \eqref{eq:leaf-ind-C2-Blocal}.
    Thus we have shown
    \[F(p)\land \project{\lnot C_3}{p} \land \project{\FlatEQ(\ell)}{p} \rightarrow I^1_{\set{p}}\land I^2_{\set{p}}.\]
\end{enumerate}

\paragraph*{Quantifier introduction and variable replacement.}
In the second part of the algorithm for the resolution rule, the variables that were supported by $\ell$ but not by the remaining literals in $C_3$ are removed from the interpolants.
We show that after each of these steps we can remove the corresponding auxiliary equation from the left-hand side of \eqref{eq:leaf-ind-with-auxeq}.
Formally, let $\varV$ be the set of auxiliary variables for the subterms of $\ell$ that are not supported by $C_3$
and that have not yet been removed in previous steps of quantifier introduction and variable replacement.
Let $\AuxEQ(\varV)$ be the auxiliary equalities for these variables and $\project{\AuxEQ(\varV)}{p}$ the conjunction of auxiliary equalities where
the head symbol of the corresponding term is in $p$ (analogously to $\project{\FlatEQ(\ell)}{p}$).
Then we show the following invariant for each step.
\begin{equation}
    F(p)\land\project{\lnot C_3}{p}\land \project{\AuxEQ(\varV)}{p} \rightarrow I^3_{\set{p}}
    \label{eq:leaf-ind-with-auxeq-var}
\end{equation}

Initially, this holds because $\varV$ is the set of all variables supported by $\ell$ but not supported by $C_3$.
Hence, $\project{\AuxEQ(\varV)}{p}$ contains all equalities of $\project{\FlatEQ(\ell)}{p}$ except those that are already
contained in $\project{\lnot C_3}{p}$.

We show that this property is preserved by any step that removes one variable by quantifier introduction or replacement.
In particular, let $v_t$ be a variable in $\varV$ corresponding to one of the outermost terms (\ie, $t$ is not a subterm of an $s$ with $v_{s} \in \varV$) and let $I^{v_t}_{\set{p}}$ be the result of 
removing the variable $v_t$ from $I^3_{\set{p}}$.  We show that after removing this variable, we have
\begin{equation*}
    F(p)\land\project{\lnot C_3}{p}\land \project{\AuxEQ(\varV\setminus\set{v_t})}{p} \rightarrow I^{v_t}_{\set{p}}.
\end{equation*}%
We show this by the three cases for $I^{v_t}_{\set{p}}$, i.e., whether $\hd(t)$ occurs only in $\set{p}$, only in $\compl{\set{p}}$, or in both.
\begin{enumerate}
    \item Case $\hd(t)$ is B-local ($\partitions(\hd(t))\cap \set{p}=\emptyset$):
    The auxiliary equality for $v_t$ is not part of the projection to $p$.
    Therefore, we already had
    \begin{align*}
        F(p) \land \project{\lnot C_3}{p} \land \project{\AuxEQ(\varV\setminus\set{v_t})}{p} \rightarrow I^3_{\set{p}}(v_t)
    \end{align*}
    The term $t$ is not supported by $C_3$ and is not a proper subterm of any term $s$  with $v_s\in\varV$.
    Hence, $v_t$ only occurs on the right-hand side and we can introduce a universal quantifier:
    \begin{align*}
        F(p) \land \project{\lnot C_3}{p} \land \project{\AuxEQ(\varV\setminus\set{v_t})}{p} \rightarrow \forall x. I^3_{\set{p}}\{ v_t \mapsto x\}.
    \end{align*}
    \item Case $\hd(t)$ is shared:
    Let $t=f(t_1,\dots,t_n)$, so $\AuxEQ(v_t)$ is $v_t = f(v_{t_1}, \dots, v_{t_n})$.
    Note that \eqref{eq:leaf-ind-with-auxeq-var} holds for all $v_t$.
    If we set $v_t = f(v_{t_1}, \dots, v_{t_n})$, then $\AuxEQ(v_t) \equiv \top$ and the implication can be simplified to
    \begin{align*}
        F(p) \land \project{\lnot C_3}{p} \land \project{\AuxEQ(\varV\setminus\set{v_t})}{p} \rightarrow I^3_{\set{p}}\{ v_t \mapsto f(v_{t_1},\dots,v_{t_n})\}.
    \end{align*}
    \item Case $\hd(t)$ is A-local:
    We start in the same way as in the second case.
    Then the right-hand side implies the existential quantified formula, and we obtain
    \begin{align*}
        F(p) \land \project{\lnot C_3}{p} \land \project{\AuxEQ(\varV\setminus\set{v_t})}{p} \rightarrow \exists x. I^3_{\set{p}}\{ v_t \mapsto x\}.
    \end{align*}
\end{enumerate}
After removing the last variable, $\varV$ is empty and \eqref{eq:leaf-ind-with-auxeq-var} implies \eqref{eq:base-ind}.

\subsection*{Tree Inductivity}  
We show that \eqref{eq:tree-ind} holds after the McMillan step and that
it is preserved by quantifier introduction and variable replacement.

\paragraph*{McMillan step.}
Let again $C_3=C_1\lor C_2$ be the resolvent for the resolution of the antecedent clauses $C_1\lor \ell$ and $C_2\lor\lnot\ell$.
Let $A_1$ and $A_2$ be sets of partitions defined by two disjoint subtrees of the tree interpolation problem
$(V, E, F )$.
For a set of partitions $A$, let $I^1_A$, $I^2_A$ and $I^3_A$ be the partial tree interpolants for $C_1$, $C_2$, and $C_3$, respectively.
We do a case distinction based on which partitions contain the
literal $\ell$.

\begin{enumerate}
    \item Case $\itpcolour(\ell)\in A_1:$ We have
    \begin{align*}
        I_{A_1}^3=I_{A_1}^1\lor I_{A_1}^2,\qquad
        I_{A_2}^3=I_{A_2}^1\land I_{A_2}^2,\qquad
        I_{A_1\dot\cup A_2}^3=I_{A_1\dot\cup A_2}^1\lor I_{A_1\dot\cup A_2}^2.
    \end{align*}
    We show \eqref{eq:tree-ind}, which with the above given partial interpolants is
    \begin{align*}
        (I_{A_1}^1\lor I_{A_1}^2) \land (I_{A_2}^1\land I_{A_2}^2) \rightarrow I_{A_1\dot\cup A_2}^1\lor I_{A_1\dot\cup A_2}^2
    \end{align*}
    When the premise is satisfied, $I_{A_1}^1\land I_{A_2}^1$ or $I_{A_1}^2\land I_{A_2}^2$ must hold. Then, by the \eqref{eq:tree-ind} property of $I^1$ and $I^2$, we know that $I_{A_1\dot\cup A_2}^1$ or $I_{A_1\dot\cup A_2}^2$ must hold as well.
    
    \item Case $\itpcolour(\ell)\in A_2:$
    \begin{align*}
        I_{A_1}^3=I_{A_1}^1\land I_{A_1}^2,\qquad
        I_{A_2}^3=I_{A_2}^1\lor I_{A_2}^2,\qquad
        I_{A_1\dot\cup A_2}^3=I_{A_1\dot\cup A_2}^1\lor I_{A_1\dot\cup A_2}^2
    \end{align*}
    The proof is symmetrical to the first case.
    \item Case $\itpcolour(\ell)\in \compl{(A_1\dot\cup A_2)}:$
    \begin{align*}
        I_{A_1}^3=I_{A_1}^1\land I_{A_1}^2,\qquad
        I_{A_2}^3=I_{A_2}^1\land I_{A_2}^2,\qquad
        I_{A_1\dot\cup A_2}^3=I_{A_1\dot\cup A_2}^1\land I_{A_1\dot\cup A_2}^2
    \end{align*}
    The proof of \eqref{eq:tree-ind} is similar to the first case. We show 
    \begin{align*}
        (I_{A_1}^1\land I_{A_1}^2) \land (I_{A_2}^1\land I_{A_2}^2) \rightarrow I_{A_1\dot\cup A_2}^1\land I_{A_1\dot\cup A_2}^2.
    \end{align*}
    By the \eqref{eq:tree-ind} property of $I^1$ and $I^2$, this implication is trivially satisfied.
\end{enumerate}

\paragraph*{Quantifier introduction and variable replacement.}

Let $I_{A_1}, I_{A_2}, I_{A_1\dot\cup A_2}$ be interpolants containing a variable $v_t$ that is removed by one step of the resolution rule.  Let $t=f(t_1,\dots,t_n)$ be the corresponding term.
We assume that \eqref{eq:tree-ind} holds before this step, \ie, 
\begin{align}
I_{A_1}(v_t) \land I_{A_2}(v_t) \rightarrow I_{A_1 \dot\cup A_2}(v_t)
\label{eq:tree-ind-with-vt}
\end{align}
and since $v_t$ occurs free it holds for every value of $v_t$.
We show that \eqref{eq:tree-ind} holds again after the variable is removed.

We do a case distinction on whether $f$ occurs in $A_1$, in $A_2$, or in $\compl{(A_1\dot\cup A_2)}$ and any Boolean combinations of these.
\begin{align}
    \partitions(f) \cap A_1 \neq\emptyset\tag{in-A1}\label{eq:hd-in-A1}\\
    \partitions(f) \cap A_2 \neq\emptyset\tag{in-A2}\label{eq:hd-in-A2}\\
    \partitions(f) \cap \compl{(A_1\dot\cup A_2)} \neq\emptyset\tag{in-B}\label{eq:hd-in-A1A2}
\end{align}
In total, we distinguish seven cases (the case where none of the above holds is not possible). For better readability, we refer to the different cases by the properties that hold, \eg, ``\emph{Case \eqref{eq:hd-in-A1}, \eqref{eq:hd-in-A2}}'' refers to the case where the first two properties hold and the third does not.

\begin{enumerate}
    \item Case \eqref{eq:hd-in-A1}: 
    The unsupported variable $v_t$ maps to a $B$-local term in $A_2$, we hence introduce a universal quantifier in front of $I_{A_2}$. The variable maps to an $A$-local term in $A_1$ and $A_1\dot\cup A_2$.
    Therefore, we introduce an existential quantifier in front of the remaining two interpolants: 
    \begin{align*}
        (\exists x_1. I_{A_1}(x_1)) \land (\forall x_2. I_{A_2}(x_2)) \rightarrow (\exists x_3.I_{A_1\dot\cup A_2}(x_3)) 
    \end{align*}

    \item Case \eqref{eq:hd-in-A2}: 
    This case is symmetric to the first case.  
    \begin{align*}
        (\forall x_1.I_{A_1}(x_1)) \land (\exists x_2.I_{A_2}(x_2)) \rightarrow (\exists x_3.I_{A_1\dot\cup A_2}(x_3)) 
    \end{align*}

    \item Case \eqref{eq:hd-in-A1A2}:
        The unsupported variable $v_t$ maps to a $B$-local term in $A_1, A_2,$ and $A_1\dot\cup A_2$. Therefore, we introduce a universal quantifier in front of all three interpolants: 
    \begin{align*}
        (\forall x_1.I_{A_1}(x_1)) \land (\forall x_2. I_{A_2}(x_2)) \rightarrow (\forall x_3.I_{A_1\dot\cup A_2}(x_3)).
    \end{align*}
    
    \item Case \eqref{eq:hd-in-A1}, \eqref{eq:hd-in-A2}:
    The unsupported variable $v_t$ maps to a shared term in $A_1, A_2$ and is replaced by its definition in the corresponding interpolants. In $A_1\dot\cup A_2$, the variable maps to an $A$-local term such that we introduce an existential quantifier in front of the corresponding interpolant.
    \begin{align*}
        I_{A_1}(f(v_{t_1},\dots,v_{t_n})) \land I_{A_2}(f(v_{t_1},\dots,v_{t_n})) \rightarrow (\exists x_3. I_{A_1\dot\cup A_2}(x_3)) 
    \end{align*}

    \item Case \eqref{eq:hd-in-A1}, \eqref{eq:hd-in-A1A2}: 
    The unsupported variable $v_t$ maps to a $B$-local term in $A_2$, we hence introduce a universal quantifier in front of $I_{A_2}$. In $A_1$ and $A_1\dot\cup A_2$, the variable $v_t$ maps to a shared term. We hence replace it by its definition in the corresponding interpolants.    
    \begin{align*}
        I_{A_1}(f(v_{t_1},\dots,v_{t_n})) \land (\forall x_2.I_{A_2}(x_2)) \rightarrow I_{A_1\dot\cup A_2}(f(v_{t_1},\dots,v_{t_n})) 
    \end{align*}

    \item Case \eqref{eq:hd-in-A2}, \eqref{eq:hd-in-A1A2}:
    This case is symmetric to the fifth case. 
    \begin{align*}
        (\forall x_1. I_{A_1}(x_1)) \land I_{A_2}(f(v_{t_1},\dots,v_{t_n})) \rightarrow I_{A_1\dot\cup A_2}(f(v_{t_1},\dots,v_{t_n})) 
    \end{align*}
    
    \item Case \eqref{eq:hd-in-A1}, \eqref{eq:hd-in-A2}, \eqref{eq:hd-in-A1A2}: 
    The unsupported variable $v_t$ maps to a shared term in all three partitions. We hence replace it by its definition in all three interpolants.
    \begin{align*}
        I_{A_1}(f(v_{t_1},\dots,v_{t_n})) \land I_{A_2}(f(v_{t_1},\dots,v_{t_n})) \rightarrow I_{A_1\dot\cup A_2}(f(v_{t_1},\dots,v_{t_n})) 
    \end{align*}
\end{enumerate}
All seven formulas are a logical consequence of \eqref{eq:tree-ind-with-vt}.
This can be shown by proving unsatisfiability of the negation of the respective implication. 
In NNF, the negated formulas in cases 1--3 contain exactly one existentially quantified formula. We use its Skolem constant to instantiate all quantified formulas and \eqref{eq:tree-ind-with-vt} to get the contradiction.
For cases 4--7, there is no existentially quantified formula.  Instantiate the universally quantified formulas and \eqref{eq:tree-ind-with-vt} with $f(v_{t_1},\dots,v_{t_n})$ to get the same contradiction.
This concludes the proof.

    \section{Bizarre Colouring of Literals}

We show here that our interpolation method also works when literals are coloured in an unusual way.
Consider the following (binary) interpolation problem.
\begin{align*}
    A:\ f(a) \neq f(s) \land a=t
    \hspace{2cm}
    B:\ t=b \land b=s
\end{align*}
Let (inA1), (inA2), (inB1) and (inB2) denote the input clauses from $A$ and $B$ in the above order.
Assume that the solver used the following lemmas to prove unsatisfiability of $A \land B$.
\begin{align}
    &a \neq t \lor t \neq b \lor a=b \tag{trans1}\label{eq:sc_trans1}\\
    &a \neq b \lor f(a)=f(b) \tag{cong1}\label{eq:sc_cong1}\\
    &f(a) \neq f(b) \lor f(b) \neq f(s) \lor f(a)=f(s) \tag{trans2}\label{eq:sc_trans2}\\
    &b \neq s \lor f(b)=f(s) \tag{cong2}\label{eq:sc_cong2}
\end{align}
Assume that the proof has the following structure.
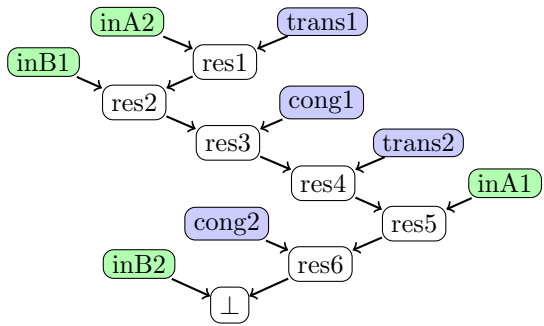
\ignore{
    \begin{prooftree}
        \def\defaultHypSeparation{\hskip 0cm}.
        \AxiomC{(cong2)}
        \AxiomC{(trans2)}
        \AxiomC{(cong1)}
        \AxiomC{(trans1)}
        \AxiomC{(inA2)}
        \BinaryInfC{(res1)}
        \AxiomC{(inB1)}
        \BinaryInfC{(res2)}
        \BinaryInfC{(res3)}
        \BinaryInfC{(res4)}
        \AxiomC{(inA1)}
        \BinaryInfC{(res5)}
        \BinaryInfC{(res6)}
        \AxiomC{(inB2)}
        \BinaryInfC{$\bot$}
    \end{prooftree}
}

\begin{center}
\begin{tikzpicture}[node distance=0.3cm and 1.5cm, align=center, every node/.style={draw, rectangle, rounded corners}]
    \node[input] (inA2) {inA2};
    \node[lemma, right=of inA2] (trans1) {trans1};
    \node[below= of {$(inA2) !.5! (trans1)$}] (res1) {res1};
    \node[input,left=of res1] (inB1) {inB1};
    \node[below=of {$(inB1) !.5! (res1)$}] (res2) {res2};
    \node[lemma, right=of res2] (cong1) {cong1};
    \node[below=of {$(res2) !.5! (cong1)$}] (res3) {res3};
    \node[lemma, right=of res3] (trans2) {trans2};
    \node[below=of {$(res3) !.5! (trans2)$}] (res4) {res4};
    \node[input, right=of res4] (inA1) {inA1};
    \node[below=of {$(res4) !.5! (inA1)$}] (res5) {res5};
    \node[lemma, left=of res5] (cong2) {cong2};
    \node[below=of {$(cong2) !.5! (res5)$}] (res6) {res6};
    \node[input, left=of res6] (inB2) {inB2};
    \node[below=of {$(inB2) !.5! (res6)$}] (bot) {$\bot$};

    \draw[->, thick] (inA2) to (res1);
    \draw[->, thick] (trans1) to (res1);
    \draw[->, thick] (inB1) to (res2);
    \draw[->, thick] (res1) to (res2);
    \draw[->, thick] (cong1) to (res3);
    \draw[->, thick] (res2) to (res3);
    \draw[->, thick] (trans2) to (res4);
    \draw[->, thick] (res3) to (res4);
    \draw[->, thick] (inA1) to (res5);
    \draw[->, thick] (res4) to (res5);
    \draw[->, thick] (cong2) to (res6);
    \draw[->, thick] (res5) to (res6);
    \draw[->, thick] (inB2) to (bot);
    \draw[->, thick] (res6) to (bot);
\end{tikzpicture}
\end{center}
Assume now that the literals appearing in $A$ (or $B$) were assigned to $A$ (or $B$), but that the literals $a = b$, $f(a) = f(b)$, and $f(b) = f(s)$ were assigned to $B$, despite $f$ being $A$-local.

The partial interpolant for \eqref{eq:sc_trans1} is $v_a = v_t$, which after resolution with the input clauses $a=t$ and $t=b$ yields $v_a = t$ for (res2) as $v_t$ becomes unsupported.
The projections of the conflict corresponding to \eqref{eq:sc_cong1} are as follows.
\begin{align*}
    &\project{\lnot C}{A}:\ v_{fa} = f(v_a) \land v_{fb} = f(v_b) \land v_a = a\\
    &\project{\lnot C}{B}:\ v_a = v_b \land v_{fa} \neq v_{fb} \land v_b = b
\end{align*}

We have to choose $p_f=A$ since $A$ is the only partition in which $f$ occurs, \ie, $\partitions(f)=\set{A}$.
The partial interpolant is hence
\begin{equation*}
    \lnot (\minproject{\lnot C}{B}) \equiv v_a \neq v_b \lor v_{fa} = v_{fb}
\end{equation*}
and therefore the partial interpolant for (res3) is $v_a = t \land (v_a \neq v_b \lor v_{fa} = v_{fb})$.
The lemma \eqref{eq:sc_trans2} yields $v_{fa} \neq v_{fs}$, and the partial interpolant for (res4) is the conjunction with the previous formula.
After resolution with input clause $f(a) \neq f(s)$, the variables $v_a$ and $v_{fa}$ become unsupported and have to be existentially quantified.
The partial interpolant for (res5) is
\begin{equation*}
    \exists v_a.\ \exists v_{fa}.\ v_a = t \land v_{fa} \neq v_{fs} \land (v_a \neq v_b \lor v_{fa} = v_{fb})
\end{equation*} or, using destructive equality reasoning and distributing $\exists$ over disjunction,
\begin{equation*}
    v_{fb} \neq v_{fs} \lor t \neq v_b \land \exists v_{fa}.\ v_{fa} \neq v_{fs}.
\end{equation*}
For the lemma \eqref{eq:sc_cong2}, we again have to choose $p_f=A$ which yields $v_b \neq v_s \lor v_{fb} = v_{fs}$ as partial interpolant.
The partial interpolant for (res6) is obtained as the conjunction with the previous interpolant, where the unsupported variables $v_{fs}$ and $v_{fb}$ are existentially quantified.
Finally, after resolution with input clause $b=s$, the variables $v_b$ and $v_s$ become unsupported. Since $s$ is shared, we replace $v_s$ by its definition. The variable $v_b$ must be universally quantified, since $b$ is $B$-local.
The final interpolant is the following formula.
\begin{equation*}
    I \equiv
    \forall v_b.\
    \exists v_{fs}, v_{fb}.\
    (v_b \neq s \lor v_{fb} = v_{fs})
    \land
    (v_{fb} \neq v_{fs} \lor t \neq v_b \land \exists v_{fa}.\ v_{fa} \neq v_{fs})
\end{equation*}

The interpolant can be simplified as follows.
\begin{align*}
    &\forall v_b.\
    \exists v_{fs}, v_{fb}.\
    (v_b \neq s \lor v_{fb} = v_{fs})
    \land
    (v_{fb} \neq v_{fs} \lor t \neq v_b \land \exists v_{fa}.\ v_{fa} \neq v_{fs})\\
    \equiv{}
    &\forall v_b.\
    \exists v_{fs}, v_{fb}.\
    (v_b \neq s \land v_{fb} \neq v_{fs})
    \lor
    (v_b \neq s \land t \neq v_b \land \exists v_{fa}.\ v_{fa} \neq v_{fs})\\
    &\hphantom{\forall v_b.\ \exists v_{fs}, v_{fb}.}
    \lor
    (v_{fb} = v_{fs} \land t \neq v_b \land \exists v_{fa}.\ v_{fa} \neq v_{fs})\\
    \equiv{}
    &\forall v_b.\
    (\exists v_{fs}, v_{fb}.\ v_b \neq s \land v_{fb} \neq v_{fs})\\
    &\hphantom{\forall v_b.\ }
    \lor
    (\exists v_{fs}, v_{fb}.\ v_b \neq s \land t \neq v_b \land \exists v_{fa}.\ v_{fa} \neq v_{fs})\\
    &\hphantom{\forall v_b.\ }
    \lor
    (\exists v_{fs}, v_{fb}.\ v_{fb} = v_{fs} \land t \neq v_b \land \exists v_{fa}.\ v_{fa} \neq v_{fs})\\
    \equiv{}
    &\forall v_b.\
    (v_b \neq s \land \exists v_{fs}, v_{fb}.\ v_{fb} \neq v_{fs})
    \lor
    (v_b \neq s \land t \neq v_b \land \exists v_{fs}, v_{fa}.\ v_{fa} \neq v_{fs})\\
    &\hphantom{\forall v_b.\ \exists v_{fs}, v_{fb}.}
    \lor
    (t \neq v_b \land \exists v_{fs}, v_{fa}.\ v_{fa} \neq v_{fs})\\
    \equiv{}
    &\forall v_b.\
    (v_b \neq s \lor (v_b \neq s \land t \neq v_b) \lor t \neq v_b)
    \land \exists v_{f1}, v_{f2}.\ v_{f1} \neq v_{f2}\\
    \equiv{}
    &(\forall v_b.\ v_b \neq s \lor (v_b \neq s \land t \neq v_b) \lor t \neq v_b)
    \land \exists v_{f1}, v_{f2}.\ v_{f1} \neq v_{f2}\\
    \equiv{}
    &t \neq s \land \exists v_{f1}, v_{f2}.\ v_{f1} \neq v_{f2}
\end{align*}
We have used that $\exists$ distributes over disjunctions in the second step, and $\forall$ over conjunctions in the fourth step, and destructive equality reasoning for $\exists$ in the third step and for $\forall$ in the last step.
Additionally, we have moved quantifiers inside formulas over subformulas where the variable did not occur.
If all terms have the same sort, the formula above simplifies to the expected interpolant
\begin{equation*}
    I \equiv t \neq s
\end{equation*}
as this disequality implies the existentially quantified formula.

\end{appendix}

\bibliographystyle{splncs04}
\bibliography{biblio}

\end{document}